\documentclass[aps,twocolumn,groupedaddress,amsmath]{revtex4}

\usepackage[dvips]{graphicx}
\usepackage{dcolumn}
\usepackage[usenames]{color}
\usepackage{bm}

\newcommand{\be}{\begin{equation}}
\newcommand{\ee}{\end{equation}}
\newcommand{\bea}{\begin{eqnarray}}
\newcommand{\eea}{\end{eqnarray}}
\newcommand{\bean}{\begin{eqnarray*}}
\newcommand{\eean}{\end{eqnarray*}}

\begin{document}


\title{Correlated geminal wave function for  molecules: an
efficient resonating valence bond approach}

\author{Michele Casula}
\email[]{casula@sissa.it}
\author{Claudio Attaccalite}
\email[]{claudio@freescience.info}
\author{Sandro Sorella}
\email[]{sorella@sissa.it}
\affiliation{International School for Advanced Studies (SISSA) Via Beirut 2,4
  34014 Trieste , Italy and INFM Democritos National Simulation Center,
  Trieste, Italy} 

\date{\today}

\begin{abstract}
We  show that a simple correlated wave function, obtained by 
applying a Jastrow correlation term to an Antisymmetrized Geminal Power
(AGP), based upon singlet pairs between electrons, is particularly suited for 
describing the electronic structure of molecules, yielding a large amount 
of the correlation energy. 
The remarkable feature of this approach is that, 
in principle, several Resonating Valence Bonds (RVB) 
can be dealt simultaneously  with a single determinant, 
at a computational cost growing with the number of electrons 
similarly to more conventional methods, such as  
Hartree-Fock (HF) or Density Functional Theory (DFT). 
Moreover  we describe an extension of 
the Stochastic Reconfiguration (SR) method, that was  recently  introduced 
for the energy minimization  
of  simple atomic wave functions. Within this extension the  
atomic positions can be considered   
as further variational parameters, that can be optimized 
together with the remaining ones.
The method is applied to several molecules from $Li_2$ to benzene by 
obtaining total
energies, bond lengths and binding energies comparable with much more
demanding multi configuration schemes.
\end{abstract}

\maketitle

\section{Introduction}

The comprehension of the nature of the chemical bond deeply
lies on quantum mechanics; since the seminal work 
by Heitler and London \cite{heitler}, 
very large steps have been made towards the possibility
to predict the quantitative properties of the
chemical compounds
from a theoretical point of view. Mean field theories, such as HF  
have been
successfully applied to a wide variety of interesting systems, although they
fail in describing those in which the correlation is crucial to characterize
correctly the chemical bonds. 
 For instance the molecular hydrogen $H_2$, 
the simplest and first studied molecule, is poorly described
by a single Slater determinant in the large distance regime, 
which is the paradigm of a strongly correlated bond; indeed, 
in order to avoid expensive energy contributions - the so called ionic terms -
that arise from two electrons 
of opposite spin surrounding the same hydrogen atom, 
one needs at least two Slater determinants to deal with a spin singlet 
wave function containing bonding and antibonding molecular orbitals.
Moreover at the bond distance it turns out that the 
resonance between those two orbitals is 
important to yield accurate bond length 
and binding energy, as the correct rate between the ionic and covalent character
is recovered. Another route that leads to the same result is to deal with an
AGP wave function, which includes the correlation in the geminal expansion;
Barbiellini in Ref.~\onlinecite{barbiellini} 
gave an illuminating example of the beauty of
this approach solving merely the simple problem of the $H_2$ molecule.

On the other hand the variational methods based on the Configuration 
Interaction (CI) technique,
which is able to take into account many Slater determinants,    
have been shown to be successful for small molecules (e.g. $Be_2$
\cite{evangelisti}). In these cases it is indeed feasible 
to enlarge the variational basis up to the saturation,
the electron correlation properties are well described and
consequently all the chemical properties can be predicted with accuracy.
However, for interesting systems with a large number of atoms 
this approach is impossible with a reasonable computational time.
Coming back to the $H_2$ paradigm,
it is straightforward to show that a gas with N $H_2$ molecules, in the dilute
limit, can be dealt accurately only with $2^N$ Slater 
determinants, otherwise one is missing important correlations due to the 
antibonding molecular orbital contributions, referred to {\em each}  
of the N $H_2$ molecules. Therefore, if the accuracy in the total energy per
atom is kept fixed, a CI-like approach does not scale polynomially 
with the number of atoms.
Although the polynomial cost of these Quantum Chemistry 
algorithms - ranging from $N^5$ to $N^7$ - is not prohibitive,  
a loss of accuracy, decreasing exponentially with the number of 
atoms is always implied, at least in their simplest variational formulations.
This is related to the loss of size consistency of a truncated CI expansion.
On the other hand, this problem can be overcome by 
coupled cluster methods, that however in their practical realization 
are not variational\cite{libro}. 

An alternative approach, not limited to small molecules, 
is based on  DFT. This theory  is in principle exact, but 
its practical implementation requires an approximation for the exchange 
and correlation functionals based on first principles, like the Local Density
approximation (LDA) and its further gradient corrections (GGA), or on
semi empirical approaches, like BLYP and B3LYP.
For this reason, even though much  effort has been made so far to go beyond  
the standard functionals, DFT is not completely reliable
in those cases in which the correlation plays a crucial role. 
Indeed it fails in describing HTc superconductors and
Mott insulators, and in predicting some transition metal 
compounds properties, whenever the outermost atomic 
d-shell is near-half-filled,
as for instance in the high potential iron proteins (HiPIP)\cite{ferro}.
Also $H_2$ molecule in
the large distance regime must be included in that list, 
since the large distance Born-Oppenheimer energy 
surface, depending on Van der Waals forces, is not well reproduced  
by the standard functionals, 
 although recently some progress 
has been made to include these important contributions\cite{vdW}.

Quantum Monte Carlo (QMC) methods are alternative to the previous ones and
until now they have been mainly used in two versions: 
\begin{itemize}
\item
Variational Monte Carlo (VMC) applied to a wave function with a Jastrow
factor that fulfills the cusp conditions and optimizes 
the convergence of the CI basis\cite{fahy,umrigarint};
\item
Diffusion Monte Carlo (DMC) algorithm  
used to improve, often by a large amount, the correlation energy of any 
given variational guess in an automatic manner \cite{dmc}. 
\end{itemize}


Hereafter we want to show that a large amount of the correlation energy
can be obtained with a single determinant, using a
size-consistent AGP-Jastrow (JAGP) wave function.
Clearly our method is approximate 
and in some cases not yet  satisfactory, 
but in a large number of interesting molecules we obtain  
results comparable and even better than multi determinants schemes based on 
few  Slater determinants per atom that are affordable by QMC only for rather 
small molecules.

Moreover, we have extended  the 
standard SR method to treat  the atomic positions as further variational  
parameters. This improvement, together with the possibility to work with 
a single determinant, has allowed us to perform a structural optimization 
in a non trivial molecule like the benzene radical cation, reaching the
chemical accuracy  with an all-electron and feasible
variational approach. 

The paper is organized as follows:
In Sec.~\ref{sec2}  we introduce the variational wave function, 
that is expanded  
over  a set of  non orthogonal atomic orbitals  
both in the  determinantal AGP and the Jastrow part.
This basis set  is consistently  optimized using 
the method described in Sec.~\ref{sec3} that, as mentioned before, 
allows also the geometry optimization.
Results and discussions  are presented in the remaining sections.

\section{functional form of the wave function}
\label{sec2}

In this paper we are going to extend the application of the JAGP wave
function, already used to study some atomic systems \cite{casula}. We
generalize its functional form in order to describe the electronic structure
of a generic cluster containing several nuclei.
With the aim to determine  a  variational  wave function, 
suitable  for a complex electronic 
system, it is important to  satisfy, as we require in the 
forthcoming chapters,  the  ''size consistency''  property: 
if we smoothly increase the distance between two regions $A$ and $B$ 
each containing  a given number  of atoms, the many-electron  
wave function $\Psi$   factorizes into the product of space-disjoint  
terms  $ \Psi = \Psi_A \bigotimes \Psi_B$ 
as long as  the interaction between the electrons coupling  the 
different regions $A$ and $B$ 
 can be neglected.  In this limit  the total   energy 
of the wave function 
approaches the sum of the energies corresponding to 
 the two  space-disjoint  regions. This property, that is obviously valid 
for  the exact many-electron ground state, is not always fulfilled by 
a generic variational wave function.
 
Our variational wave 
function is defined by the product of two terms, namely  
a Jastrow $J$ and 
an antisymmetric part ($\Psi=J \Psi_{AGP}$).
If the former  
is an explicit contribution to the dynamic 
electronic correlation, the latter  is able
to treat the non dynamic  one  
arising  from near degenerate orbitals 
 through the geminal expansion. 
Therefore our wave
function is highly correlated and it is expected to give accurate results
especially  for molecular systems.
The Jastrow term is further split into  a two
 body and a three body factors, $J = J_2 J_3$, described 
in the following subsections after the AGP part.

\subsection{Pairing determinant}

 As is well known, a simple Slater determinant 
provides the exact exchange electron interaction  but neglects the
electronic correlation, which is by definition 
the missing energy contribution. 
In the past different strategies were proposed to go beyond Hartree-Fock 
theory.
In particular a sizable amount of the correlation energy is 
obtained by applying to a Slater determinant a so called Jastrow term,  
 that explicitly takes into account the pairwise 
interaction between electrons.  
QMC allows to deal with this term in an efficient way\cite{foulkes}.
On the other hand, within the 
Quantum Chemistry community AGP is a well known ansatz to improve 
the HF theory, because  it implicitly includes most of the 
double-excitations of an HF state. 
  
Recently we proposed a new trial function for atoms, 
that includes both the terms. Only the interplay between them 
yields in some cases, like Be or Mg,  
an  extremely accurate description of the correlation energy.
In this work we extend  this promising approach to  a number 
of small molecular systems with known experimental 
 properties, that are  commonly used 
for testing  new numerical  techniques.  
 
 The major advantage of this
approach is the inclusion of many CI expansion terms with the computational
cost of a single determinant, 

that allow us to extend the calculation with 
a full structural optimization up to benzene, without a 
particularly heavy computational effort on a single processor machine. 
For an unpolarized system containing $N$ electrons
(the first $N/2$ coordinates are referred to the up spin electrons)  
the AGP wave function is a $\frac{N}{2} \times \frac{N}{2}$ pairing matrix
determinant, which reads:  
\be
\Psi_{AGP}(\textbf{r}_1,...,\textbf{r}_N) =
\det \left (\Phi_{AGP}(\textbf{r}_i,\textbf{r}_{j+N/2}) \right ),
\ee
and the geminal function is expanded over an atomic basis:
\be
\Phi_{AGP}(\textbf{r}^\uparrow,\textbf{r}^\downarrow)
=\sum_{l,m,a,b}{\lambda^{l,m}_{a,b}\phi_{a,l}
(\textbf{r}^\uparrow)\phi_{b,m}(\textbf{r}^\downarrow)} ,  
\label{expgem}
\ee
where indices $l,m$ span different orbitals centered on atoms $a,b$, and
$i$,$j$ are coordinates of spin up and down electrons respectively.
The geminal functions may be viewed as an extension of the simple HF wavefunction, based on molecular orbitals, and in fact the geminal function coincide with HF only when the number $M$ of non zero eigenvalues of the $\lambda$ matrix is equal to $N/2$.
Indeed the general function \ref{expgem} can be written in diagonal form after an appropriate transformation:
\be
\Phi_{AGP}(\textbf{r}^\uparrow,\textbf{r}^\downarrow)
=\sum_{k}^M{\lambda^{k}\tilde{\phi}_{k}
(\textbf{r}^\uparrow)\tilde{\phi}_{k}(\textbf{r}^\downarrow)} ,  
\ee
where $\tilde{\phi}_{k}(\textbf{r}) = \sum_{j,a} \mu_{k,j,a} \phi_{j,a}(\textbf{r})$ are just the molecular orbitals of the HF theory whenever $M=N/2$. 
Notice that with respect to our previous pairing function formulation 
also off-diagonal elements are now included in the $\lambda$ matrix, 
which must 
be symmetric in order to define a singlet spin orbital state. 
Moreover it allows one to easily fulfill other system symmetries by setting the
appropriate equalities among different $\lambda_{l,m}$. 
For instance in homo-nuclear diatomic molecules, the invariance under
reflection in the plane perpendicular to the molecular axis
yields the following relation: 
\be \label{eqmich}
\lambda^{a,b}_{m,n}=(-1)^{p_m+p_n} \lambda^{b,a}_{m,n},
\ee
where $p_m$ is the parity under reflection of the $m-$th orbital.

Another important property of this formalism is the possibility to describe
resonating bonds present in many structures, like benzene. A
$\lambda^{a,b}_{m,n}$ different from zero represents a chemical bond formed by
the linear combination of the \emph{m}-th and \emph{n}-th orbitals belonging 
to \emph{a}-th and \emph{b}-th nuclei. 
It turns out that resonating bonds can be well described through the geminal
expansion switching on the appropriate $\lambda^{a,b}_{m,n}$ coefficients: the
relative weight of each bond is related to the amplitude of its $\lambda$.   

Also the spin polarized molecules can
be treated within this framework, by using the spin generalized version of the
AGP (GAGP), in which the unpaired orbitals are expanded as well as the 
paired ones over the same
atomic basis employed in the geminal\cite{coleman}.
As already mentioned in the introduction of this chapter, the size consistency
is an appealing feature of the AGP term. Strictly  speaking, the AGP wave
function is certainly  size consistent when 
both the compound and the separated fragments have the minimum 
possible total spin,
 because the geminal expansion contains both bonding and
antibonding contributions, that can mutually cancel the ionic term arising in
the asymptotically separate regime. 
Moreover  the size consistency of the AGP, as well 
as  the  one of the Hartree-Fock state,  holds in all cases in which 
the spin of the compound is the sum of the spin of the fragments.
 However, similarly to other approaches\cite{libro}, 
  for spin polarized systems the size consistency 
 does not generally hold,  
and, in such cases, 
 it may be important  go beyond a
single AGP wave function. 
Nevertheless  we have experienced that a  single reference AGP
state is able to describe accurately the electronic structure of the compound 
around the Born-Oppenheimer minimum even in the mentioned 
polarized cases, such as in the oxygen dimer.

The last part of this section is devoted to the nuclear cusp condition
implementation.
A straightforward calculation shows that the AGP wave function fulfills the
cusp conditions around the nucleus $a$ if the following linear system is
satisfied: 
\be
\sum^{(1s,2s)}_{j}{\lambda^{j,j^\prime}_{a,b}
  \hat \phi'_{a,j}(\textbf{r}=\textbf{R}_a)}=-Z_a
\sum_{c,j}{\lambda^{j,j^\prime}_{c,b}\phi_{c,j} 
(\textbf{r}=\textbf{R}_a)} ,
\label{cuspcond}
\ee
for all $b$ and $j^\prime$; in the LHS the caret denotes the spherical average
of the orbital gradient. 
The system can be solved iteratively during the optimization processes, but if
we impose that the orbitals satisfy the single atomic cusp conditions, it
reduces to: 
\be
\sum_{c(\neq a),j}{\lambda^{j,j'}_{c,b}\phi_{c,j}(\textbf{R}_a)} = 0,
\label{reducecusp}
\ee
and because of the exponential orbital damping, if the nuclei are not close
 together each term in the previous equations is very small, 
of the order of $\exp (-|\textbf{R}_a-\textbf{R}_c|)$.
Therefore, with the aim of making the optimization faster, 
we have chosen to use $1s$
and $2s$ orbitals satisfying the atomic cusp conditions and to disregard the
sum (\ref{reducecusp}) in Eq.~\ref{cuspcond}. 
In this way, once the energy minimum is reached,  
also the molecular cusp conditions (\ref{cuspcond}) are rather well 
 satisfied. 

\subsection{Two body Jastrow term}
 
As it is well known the Jastrow term plays a crucial role in treating
many body correlation effects. One of the most important correlation
contributions arises from the electron-electron interaction. Therefore it is
worth using at least a two-body Jastrow factor in the trial wave
function. Indeed  this term reduces the electron coalescence probability, and
so decreases 
the average value of the repulsive interaction.  
The two-body Jastrow function  reads:
\be
J_2(\textbf{r}_1,...,\textbf{r}_N) = 
\exp{\left (\sum_{i<j}^{N}u(r_{ij}) \right )}, 
\ee
where  $u(r_{ij})$  depends only on the relative distance $r_{ij}= 
|\textbf{r}_i -\textbf{r}_j|$  between 
two electrons and allows to  fulfill the cusp conditions 
for opposite spin electrons as long as 
$u(r_{ij})\to \frac{r_{ij}}{2}$ for small  electron-electron distance. 
The pair correlation function $u$ can be parametrized successfully by 
few variational parameters. We have adopted two main functional forms. 
The first is similar to the one given by Ceperley \cite{cep2body}: 
\be \label{simfahy}
u(r) = \frac{F}{2}\left ( 1 - e^{-{r}/F}\right),
\ee
with $F$ being a free variational parameter. 
This form for $u$ 
is particularly convenient whenever atoms are very far apart at 
distances much larger than $F$, 
as it  allows to obtain  good size consistent energies,  approximately 
equal to the sum of the  atomic contributions, without changing the 
other parts  of the wave function with an expensive optimization.
Within  the functional form  (\ref{simfahy}),  
it is assumed that the long range part of the Jastrow, 
decaying as a power of the distance between atoms, is included in the 
3-body Jastrow term described in the next subsection. 
The second form of the pair function $u$, particularly convenient at 
the chemical bond distance,  where we performed most of the calculations, 
is the one used by Fahy \cite{fahy}:
\be
u(r)= { \frac{r}{2 (1 + b r )} },
\label{twobodyold}
\ee
with a different variational parameter $b$.

In both functional forms 
the  cusp condition for antiparallel spin electrons is satisfied, whereas 
the one for parallel spins is neglected in order to avoid the 
spin contamination.
This allows to  remove  the singularities of the local energy due to the
collision 
 of two opposite spin electrons, yielding a smaller variance and a more
 efficient VMC calculation. Moreover, due to the Jastrow correlation,  
an exact property of the ground state wave function is  
recovered  without using many Slater determinants, thus 
considerably simplifying  the variational parametrization of a correlated 
wave function. 

\subsection{Three Body Jastrow term}
\label{3bodysec}
In order to describe well the correlation between electrons the simple 
Jastrow factor is not sufficient. Indeed it takes into account only 
the electron-electron separation and not the individual electronic position 
$\mathbf{r}_i$ and $\mathbf{r}_j$. It is expected that close to nuclei the electron 
correlation is not accurately described by a translationally invariant  Jastrow, as shown by different
authors, see for instance Ref. \onlinecite{umrigar3body}. 
For this reason we introduce a factor, often called three body 
(electron-electron-nucleus) Jastrow,
that explicitly depends on both the electronic positions 
$\mathbf{r}_i$ and $\mathbf{r}_j$.  
The  three body Jastrow is chosen to satisfy the 
following requirements: 
\begin{itemize}
\item 
The cusp conditions set up by the two-body Jastrow term and by the AGP 
are preserved.
\item 
It does not distinguish the electronic spins 
otherwise causing  spin contamination. 
\item 
Whenever the atomic distances are large it factorizes 
into  a product of independent contributions located near each atom, 
an important  requirement  to satisfy the  
size consistency of the variational wave function.
\end{itemize}

Analogously to  the pairing trial function in Eq.~\ref{expgem} 
we define  a three body factor as:
\bea \label{3body}
J_3(\textbf{r}_1,...,\textbf{r}_N) 
&=& \exp \left( \sum_{i<j} \Phi_J(\textbf{r}_i,\textbf{r}_j) \right)
\nonumber \\
\Phi_J(\textbf{r}_i,\textbf{r}_j) &=& \sum_{l,m,a,b} g_{l,m}^{a,b}\psi_{a,l}
(\textbf{r}_i)\psi_{b,m} (\textbf{r}_j), 
\eea
where indices $l$ and $m$ 
indicate different orbitals  located around the atoms $a$ and $b$ 
respectively. 
Each Jastrow orbital $\psi_{a,l}(\textbf{r})$ is  centered on 
the corresponding  atomic position $\textbf{R}_a$. 
We have used Gaussian and exponential orbitals multiplied by appropriate 
polynomials of the electronic coordinates, related to different 
spherical harmonics with given angular momentum, 
 as in the usual Slater basis. 
Analogously to the geminal function $\Phi_{AGP}$,  whenever the one particle
basis set $\{ \psi_{a,i} \}$ is complete the  
expansion (\ref{3body}) is also complete for the generic two particle  
function $\Phi_J(\mathbf{r},\mathbf{r}^\prime)$. In the latter  case, however,
the one particle orbitals have to behave smoothly 
close to the corresponding  nuclei, namely  as:
\be
\psi_{a,i} (\mathbf{r}) -\psi_{a,i} ( \mathbf{R}_a ) 
 \simeq |\mathbf{r}-\mathbf{R}_a|^2,
\ee
or with larger power, 
in order to preserve the nuclear cusp conditions (\ref{cuspcond}).

For the s-wave orbitals we have found energetically convenient to add   
a finite constant $c_{l} / (N-1)$.
As shown in the Appendix~\ref{consistency}, 
a non zero value of the  constant $c_l$ 
for  such  orbitals  $\psi_{a,l}$ is equivalent to include 
in the wave function a size consistent one body term.
As pointed out in  Ref.~\onlinecite{krotscheck}, 
it is easier to optimize a one body term implicitly 
present in the 3-body Jastrow factor, 
 rather than including more orbitals in the determinantal basis set.  

The chosen form for the 3-body Jastrow (\ref{3body}) is similar to one used
by Prendergast et al. \cite{prendergast} and
has very appealing features: it easily allows including the
symmetries of the system by imposing them on the matrix 
$g_{l,m}^{a,b}$ exactly  as it is possible  for the 
 pairing part (e.g. by replacing $\lambda_{m,n}^{a,b}$  with 
 $g_{m,n}^{a,b}$ in Eq.~\ref{eqmich}).
 It is size consistent, namely the atomic limit  
can be smoothly recovered with the same trial function  
when the  matrix terms $g_{l,m}^{a,b}$ for $a \ne b$ 
approach zero in this limit. Notice that a small non zero value  
of $g_{l,m}^{a,b}$ for $a\ne b$ acting on p-wave orbitals 
can correctly describe 
a weak interaction between electrons such as the 
the  Van der Waals forces.

\section{Optimization method}
\label{sec3}
We have used the Stochastic Reconfiguration (SR) method 
 already described in Ref.~\onlinecite{sorella}, that allows 
to minimize the energy expectation value of a variational wave function containing many variational parameters in an arbitrary functional form.
The basic ingredient for the stochastic minimization of 
the wave function $\Psi$ determined by $p$ variational parameters 
$\{\alpha^0_k\}_{k=1,\ldots,p}$, is the solution of the linear system:
\begin{equation} \label{srstandard}
\sum\limits_{k=0}^p  s_{j,k}  \Delta  \alpha_k = \langle \Psi | O_k (\Lambda
I - H)  |\Psi \rangle ,  
\end{equation}
where the operators $O_k$ are defined on each 
$N$ electron configuration $x=\{ \mathbf{r}_1, \ldots , \mathbf{r}_N \}$ 
as the logarithmic derivatives with respect to the parameters $\alpha_k$:
\begin{equation} \label{defok}
O^k(x)= \frac{ \partial }{ \partial \alpha_k } \ln \Psi (x) 
\textrm{~~~for $k > 0$} ,
\end{equation}
while for $k=0$ $O_k$ is the identity operator  
equal to one on all the configurations.  
The $(p+1) \times (p+1)$  matrix $s_{k,j}$ is easily expressed  in terms 
of these operators:
\begin{equation}  \label{defsr}
s_{j,k}= { \frac{\langle \Psi | O_j O_k |\Psi \rangle} {\langle \Psi |\Psi
  \rangle }} ,
\end{equation}
and is calculated at each iteration through a standard variational 
Monte Carlo sampling ; the single iteration constitutes a small simulation 
that will be referred in the following as ``bin''.
After each bin the wave function parameters  
are iteratively updated ($\alpha_k \to \alpha_k + \Delta \alpha_k/ \Delta
\alpha_0 $), and the method is convergent to an energy minimum 
for large enough $\Lambda$. Of course for particularly simple functional form
of $\Psi(x)$, containing e.g. only linear CI coefficients,  much more
efficients optimization schemes do exist \cite{night}. 

SR is similar to a standard steepest descent (SD) calculation, where 
the expectation value of the  energy 
$E( \alpha_k)={ \frac{\langle  \Psi | H |  \Psi \rangle} {\langle \Psi | \Psi
  \rangle }}  $ is optimized by iteratively changing 
  the parameters $\alpha_i$ according to
 the corresponding derivatives of the energy (generalized forces):
\begin{widetext}
\begin{equation} \label{forces}
f_k = - {\partial E \over \partial  \alpha_k} 
= - { \langle \Psi | O_k H + H O_k  + (\partial_{\alpha_k} H) | \Psi \rangle 
 \over \langle \Psi | \Psi \rangle }
+ 2 {   \langle \Psi |  O_k |\Psi  \rangle \langle \Psi |  H | \Psi \rangle 
\over \langle \Psi | \Psi \rangle^2 },
\end{equation}
\end{widetext}
namely:
\begin{equation} 
 \alpha_k \to \alpha_k +   \Delta t    f_k.
\end{equation}
$\Delta t$ is a suitable small time step, which can be taken 
fixed or determined at each iteration 
 by minimizing the energy expectation value.
Indeed the variation of the total energy $\Delta E$ 
at each step is easily shown to 
be negative  for small enough $\Delta t$ because, in this 
limit $$\Delta E = - \Delta t \sum_i f_i^2 + O(\Delta t^2).$$
Thus the method certainly converges  at the  minimum when all the forces 
vanish. Notice that in the definition of the generalized forces 
(\ref{forces}) we have generally assumed 
that the variational parameters may appear also in the Hamiltonian.
This is particularly important for  the structural optimization
since the atomic positions that minimize the energy enter both in the
wave function and in the potential.
  
In the following we will show that 
similar considerations hold for the SR method, that can be therefore 
extended to the optimization of the geometry. 
Indeed, by eliminating the equation with index $k=0$ from 
the linear system (\ref{srstandard}), 
the SR iteration can be written in a form similar to the 
steepest descent:
\begin{equation} \label{iterforce}
\alpha_i \to \alpha_i +  \Delta t   \sum_k \bar s^{-1}_{i,k} f_k
\end{equation}   
where the reduced $p \times p$ matrix $\bar s$ is:
\begin{equation}
\bar s_{j,k} = s_{j,k} - s_{j,0} s_{0,k} 
\end{equation}
and the $\Delta t$ value is given by:
\begin{equation}
\Delta t= { 1 \over 2  ( \Lambda -{ \langle \Psi |H| \Psi \rangle 
\over \langle \Psi|\Psi\rangle }  
- \sum_{k>0}  \Delta \alpha_k s_{k,0} ) }.
\end{equation}
>From the latter equation 
the value of $\Delta t$ changes during the simulation 
and remains small for large enough energy shift  $\Lambda$.
However,  using the analogy with  the steepest descent, 
convergence to the energy minimum is reached also when  
the value of $\Delta t$ is sufficiently small and 
is kept constant for each iteration.
Indeed the energy variation for a small change of the parameters is:
 $$\Delta E = -\Delta t  \sum_{i,j} \bar s^{-1}_{i,j} f_i f_j.$$
It is easily verified that the above term 
 is always negative because  the reduced matrix $\bar s$,  as well 
as $\bar s^{-1}$,  is positive definite, being  
 $s$ an overlap matrix with all positive eigenvalues. 

For a stable iterative method, such as the SR or the SD one, 
a basic ingredient is that at each iteration the new parameters 
$\alpha^\prime$ are close to the previous $\alpha$ according to 
a prescribed distance. 
The fundamental difference between the SR minimization and the standard 
steepest descent is just related to the 
definition of this  distance.
For the SD it is the usual one defined by the Cartesian metric
$ \Delta_\alpha = \sum_k  | \alpha^\prime_k  - \alpha_k|^2$, instead the SR  
works correctly in the physical Hilbert space metric of the 
wave function $\Psi$, yielding 
$\Delta_\alpha=  \sum_{i,j} \bar s_{i,j}  (\alpha^\prime _i-\alpha_i) 
( \alpha^\prime_j-\alpha_j),$ namely  the square distance between 
the two normalized wave functions corresponding to the two 
 different sets of variational 
parameters $\{ \alpha^\prime \}$ and $\{ \alpha_k \}$. 
 Therefore, from the knowledge of the 
generalized forces $f_k$, the most convenient change 
of the variational parameters minimizes the functional 
$\Delta E +\bar \Lambda  \Delta_\alpha  $,
where  $\Delta E$ is the linear change in the energy 
 $\Delta E = -\sum_{i} f_i (\alpha^\prime_i-\alpha_i) $
and $\bar \Lambda$  is a Lagrange multiplier that allows a stable minimization 
with small change $\Delta_\alpha$ of the wave function $\Psi$.
 The final iteration (\ref{iterforce})   is then  
easily  obtained. 

The advantage of SR compared with SD is obvious because sometimes a small 
change of the  variational parameters correspond to a large change 
of the wave function, and the SR takes into account this effect through 
the Eq.~\ref{iterforce}. 
In particular  the method is useful when a non orthogonal basis set 
 is used as we have done in this work.  
Indeed by using the reduced matrix $\bar s$ 
it is also possible to remove from 
the calculation those parameters that imply some redundancy in 
the variational space. As shown in the Appendix~\ref{srstable}, 
a more efficient change in the wave function can be obtained by updating  
only the variational parameters that remain independent 
within a prescribed tolerance, 
and therefore, by removing the parameters that linearly depend 
on the others. A more stable minimization 
is obtained without spoiling the accuracy of the calculation.
 A weak tolerance criterium $\epsilon \simeq 10^{-3}$, 
provides a very stable algorithm even when the dimension of the variational 
space is large. 
For a small atomic basis set, by  an appropriate  choice of the 
Jastrow and Slater orbitals,
the reduced  matrix $\bar s$ is always very  well conditioned 
even for the largest system studied,  and  
the above stabilization technique is not required.  
Instead the described method is particularly important for the extension of   
QMC to complex systems with large number of atoms and/or  higher 
level  of accuracy, because in this case  it is very difficult 
to select - e.g. by trial and error - 
the relevant variational parameters, that allow 
a well conditioned matrix $\bar s$ for a stable inversion in 
(\ref{iterforce}). 

\subsection{Setting the SR parameters}  

In this work, instead of setting the constant $\Lambda$, 
we have equivalently chosen  to determine $\Delta t$
by verifying the stability and the convergence of the algorithm at 
fixed $\Delta t$ value, which can be easily understood
 as an inverse energy scale. 
The simulation is stable whenever $ 1 / \Delta t > {\rm  \Lambda_{cut} } $,
where $\Lambda_{cut}$ is an energy cutoff that is strongly 
dependent  on the chosen wave function 
and is generally weakly dependent on the bin length. 
Whenever the wave function is too much detailed, namely has a lot of variational freedom,
especially for the high energy components of the core electrons, 
the value of $\Lambda_{cut}$ becomes exceedingly large
and too many iterations are required for obtaining a 
converged variational wave function.
In fact a rough estimate of the 
corresponding number of iterations  $P$ is given by  
$P \Delta t >>  1/G$, where $G$ is the typical energy gap 
of the system, of the order of few electron Volts 
in small atoms and molecules.   
Within the SR method it is therefore extremely important to work with a 
bin length rather small, so that many iterations can be performed without 
much effort. 

\begin{figure*}[!ht]
\includegraphics[width=13cm,height=7cm]{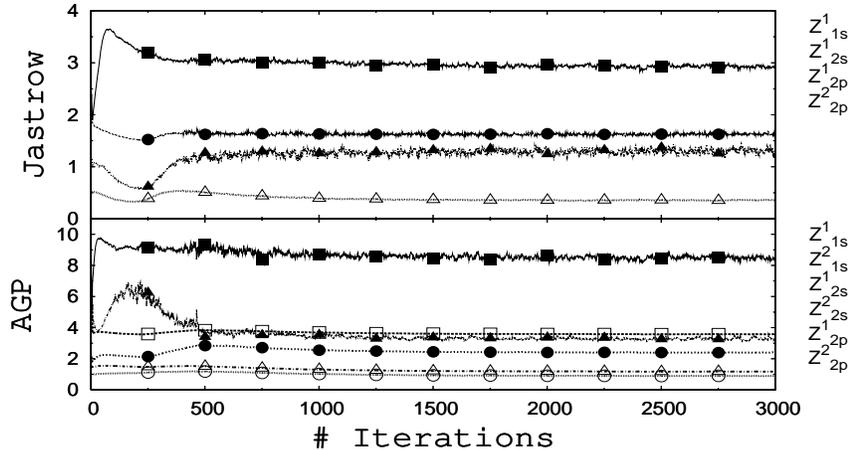}
\caption{\label{parameters}
Example of the convergence of the SR method for the  
variational parameters of the Be atom, as a function of the 
number of stochastic iterations. In  the upper(lower) panel  
the Jastrow (geminal) parameters are shown.
For each iteration, a variational Monte Carlo 
calculation is employed with a bin containing $15000$  
samples of the energy, yielding at the equilibrium 
a standard deviation of $\simeq 0.0018 H$. 
For the first 200 iteration $\Delta t= 0.00125 H^{-1} $, 
for the further 200 iterations   $\Delta t= 0.0025 H^{-1}$, 
whereas for the remaining ones $\Delta t= 0.005 H^{-1} $.}
\end{figure*}

In a Monte Carlo optimization framework the forces $f_k$ are always determined 
with some statistical noise $\eta_k$, and by iterating 
the procedure several times with a fixed bin length 
the variational parameters will fluctuate around their mean 
values.  These statistical fluctuations  are similar  to the  
thermal noise of a standard Langevin equation:
\begin{equation}
\partial_t  \alpha_k = f_k +\eta_k ,
\end{equation}
where 
\be
\langle \eta_k (t) \eta_{k^\prime} (t^\prime) \rangle= 
2  T_{noise} \delta (t-t^\prime) \delta_{k,k^\prime}.
\label{noiseforce}
\ee
The variational parameters $\alpha_k$, averaged over the 
Langevin simulation time (as for instance in Fig.\ref{parameters} for 
$t> 2 H^{-1}$),  will be close to the true energy minimum, 
but the corresponding forces
 $f_k= -{ \partial_{\alpha_k}   E } $ will be
affected by a bias that scales to zero with the thermal noise $T_{noise}$, due to the presence of non quadratic terms in the energy landscape. 

Within a  QMC  scheme, one needs to estimate $T_{noise}$, by increasing the bin length as clearly 
$T_{noise} \propto 1/ {\rm Bin~length }$, because the statistical fluctuations of the forces, obviously decreasing by increasing the bin length, are related to the thermal noise by Eq. \ref{noiseforce}. Thus there is an optimal value for the bin length, which guarantees a fast
convergence and avoid the forces to be biased within the statistical accuracy
of the sampling.

\begin{figure}[!ht]
\includegraphics[width=8.5cm]{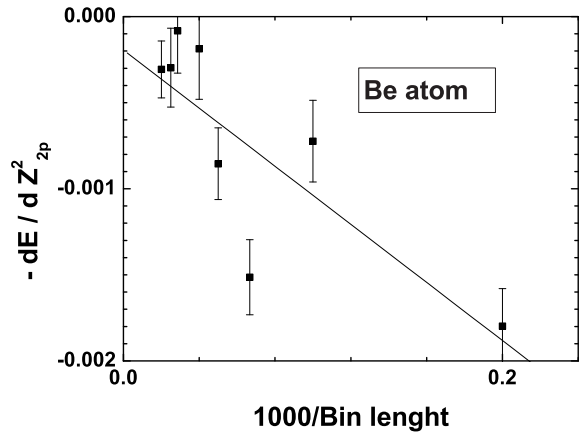}
\caption{\label{forzap19}
Calculation of the derivative of the energy with respect to the 
second $Z$ in the $2p$ orbital of the geminal function for the Be atom.
The calculation of the force was obtained, at fixed variational 
parameters,  by averaging over $10^7$ 
samples, allowing e.g. a statistical accuracy in the total energy of 
$0.07 mH$. The variational parameters have been obtained by 
an SR minimization with fixed bin length shown in the x label.
The parameter considered has 
 the largest deviation  from the Euler  conditions.}
\end{figure}

An example is shown in Fig.~\ref{parameters} for the 
optimization of the Be atom, using a DZ basis both for the geminal 
and the three-body Jastrow part.  The convergence is reached in  about 
1000 iteration with $\Delta t= 0.005 H^{-1}$. However, in this case 
it is possible to use a small bin length,  
yielding a statistical accuracy in the energy much poorer   than the 
final accuracy of about $0.05 mH$.
This is obtained 
by averaging the variational 
parameters in the last $1000$ iterations, 
when they fluctuate around a mean value,  allowing
a very accurate determination of the energy minimum which
satisfies the Euler conditions, namely with $f_k=0$ for all parameters.
Those conditions have been tested   
by an independent 
 Monte Carlo simulation about $600$ times longer  than the bin used 
during the minimization.
As shown in Fig.~\ref{forzap19} 
the Euler conditions are fulfilled within statistical accuracy 
even when the bin used for the minimization is much smaller than the 
overall simulation. On the other hand if the bin used is too small, 
as we have already pointed out, the averaging of the parameters is 
affected by a sizable bias. 

 Whenever it is possible to use a relatively small bin 
in the minimization,  the apparently large 
number of iterations required for equilibration does not really matter, 
because a comparable amount of time has to be  spent in the averaging 
of the variational parameters, as shown in Fig.~\ref{parameters}. 
The comparison shown in the Ref.~\onlinecite{filippi2} about the number of the 
iterations  required,  though is clearly relevant for a deterministic method, 
is certainly incomplete for a statistical method, because in the latter case
an iteration can be performed in principle in a very short time, namely with 
a rather  small bin.

It is indeed possible that for high  enough accuracy the number 
of iterations needed for the equilibration becomes  negligible  
from the computational point of view. In fact in order to reduce,   e.g. 
by a factor ten,  the accuracy in the variational parameters, a bin  
ten times larger is required  
for decreasing  the thermal noise $T_{noise}$ by the same factor,  
whereas a statistical average $100$ times longer is 
indeed necessary to reduce the statistical errors of the variational parameters by the same ratio.
This means that the fraction of 
time spent for  equilibration  
becomes ten times smaller  compared with the less accurate simulation.

\subsection{Structural optimization} 
\label{structural}

In the last few years remarkable progresses have been made to develop 
Quantum Monte Carlo (QMC) techniques which are able in principle 
to perform structural optimization of molecules and complex systems
\cite{zeroforce,penalty,correlated}. 
Within the Born-Oppenheimer approximation the nuclear positions 
$\mathbf{R}_i$ can be considered as further 
variational parameters included in the 
set $\{ \alpha_i \}$ used for the SR minimization (\ref{iterforce}) 
of the energy expectation value. 
For clarity, in order to distinguish the 
conventional variational parameters from the ionic positions, 
in this section we indicate
with $\{c_i\}$ the former ones, and with $\mathbf{R}_i$
the latter ones. It is understood that  $R_i^\nu =\alpha_k$,
where a particular index $k$ of the whole set of parameters $\{ \alpha_i \}$
corresponds to a given spatial component ($\nu=1,2,3$) of the $i-$th ion.
Analogously the forces (\ref{forces}) acting on the ionic positions 
will be indicated by capital letters with the same index notations.

The  purpose  of the present  section is to compute 
the  forces $\mathbf{F}$ acting on  each  of the  $M$ nuclear positions
$\{\mathbf{R}_1, \ldots , \mathbf{R}_M\}$, being $M$ the total
number of nuclei in the system:
\be
\label{force}
\mathbf{F}(\mathbf{R}_a)=-\mathbf{\nabla}_{\mathbf{R}_a}
 E(\{c_i\},\mathbf{R}_a),
\ee
with a reasonable statistical accuracy, so that the iteration 
(\ref{iterforce}) can be effective for the structural optimization. 
In this work we have used a finite difference operator $ { \mathbf{\Delta \over 
\Delta R}_a } $ for the  evaluation of the 
force acting on a given  nuclear position $a$: 
\begin{eqnarray} \label{forcefinite}
\mathbf{F} (\mathbf{R}_a) & = & - { \mathbf{\Delta \over \Delta R}_a } E 
\\ \nonumber
   & = &
- { E(\mathbf{R}_a + \mathbf{\Delta R}_a )  
 -  E(\mathbf{R}_a - \mathbf{\Delta R}_a )    
  \over 2 \Delta R   } + O({\Delta R}^2) 
\end{eqnarray}
where $\mathbf{\Delta R}_a$ is a 3 dimensional vector.
Its length $\Delta R $ is chosen to be $0.01$ atomic 
units, a value that is small enough for negligible finite 
difference errors.
In order to evaluate the energy differences in Eq.~\ref{forcefinite} 
we have used the space-warp coordinate transformation \cite{umrigar,filippi} 
briefly summarized  in the following paragraphs.
According to this   transformation also the electronic coordinates 
$\mathbf{r}$ will be translated in order to mimic
the right displacement of the charge around the nucleus  
$a$:  
\be \label{spacewarp}
\overline{\mathbf{r}}_i=\mathbf{r}_i+
\mathbf{\Delta R}_a ~ \omega_a(\mathbf{r}_i),
\ee
where
\be
\omega_a(\mathbf{r})=\frac{F(|\mathbf{r}-\mathbf{R}_a|)}
{\sum_{b=1}^{M} F(|\mathbf{r}-\mathbf{R}_b|)}.
\ee
$F(r)$ is a function which must decay rapidly; here we used 
$F(r)=\frac{1}{r^4}$ as suggested in Ref.~\onlinecite{filippi}.

The expectation value of the energy depends on
$\mathbf{\Delta R}$, because both the Hamiltonian and the wave function depend
on the nuclear positions. Now let us apply the space-warp transformation to
the integral involved in the calculation; the expectation value reads:
\be \label{forcewarp}
E( \mathbf{R}  +\mathbf{\Delta R})=
\frac{\int d\mathbf{r} J_{\mathbf{\Delta R}}(\mathbf{r})
\Psi_{\mathbf{\Delta R}}^2 (\overline{\mathbf{r}}(\mathbf{r}))
E^{\mathbf{\Delta R}}_L(\overline{\mathbf{r}}(\mathbf{r}))}
{\int d\mathbf{r} J_{\mathbf{\Delta R}}(\mathbf{r}) 
\Psi^2_{\mathbf{\Delta R}}(\overline{\mathbf{r}}(\mathbf{r}))},  
\ee
where $J$ is the Jacobian of the transformation and here and henceforth we 
avoid for simplicity to use the atomic subindex $a$.
The importance of the space warp in reducing the 
variance of the force is easily understood for the case of an isolated 
atom $a$. Here the force acting on the atom is obviously 
 zero, but only after the space warp transformation 
with $\omega_a=1$ the integrand  
of expression (\ref{forcewarp}) will be independent of $\mathbf{\Delta R}$, 
providing an estimator of the force with zero variance. 

Starting from Eq.~\ref{forcewarp}, 
it is straightforward to explicitly derive a finite difference 
differential expression for the force estimator, 
which is related to the gradient of the previous quantity with respect
to $\mathbf{\Delta R}$, in the limit of the displacement tending to zero:
\begin{widetext}
\begin{equation}
\label{vmcforce}
\mathbf{F}(\mathbf{R}) 
  =  - \big \langle  \lim_{\mathbf{|\Delta R|} \rightarrow 0}
\mathbf{ \frac{\Delta}{\Delta R}} E_L \big \rangle  
 +   2 \Big ( 
\big \langle H \big \rangle \big \langle\lim_{\mathbf{|\Delta R|} \rightarrow 0}
\mathbf{ \frac{\Delta}{\Delta R}}
\log (J^{1/2}  \Psi  ) \big \rangle -
\big \langle H \lim_{\mathbf{|\Delta R|} \rightarrow 0}  
\mathbf{ \frac{\Delta }{ \Delta R}}
\log (J^{1/2}  \Psi  ) \big \rangle 
\Big ),
\end{equation}
\end{widetext}
where the brackets indicate a Monte Carlo like average over the 
square modulus of the trial wave function, 
$ \mathbf{ \Delta \over \Delta R }$ is the finite difference 
derivative as defined in (\ref{forcefinite}), and $E_L= 
{ \langle \Psi |H |x \rangle \over \langle \Psi |x \rangle } $ is 
 the local energy on a configuration $x$ where all electron positions 
and spins are given. 
In analogy with the general expression (\ref{forces}) of the 
forces, we can identify 
the operators 
 $O_k$  corresponding to the space-warp 
change of the variational wave function:
\begin{equation} \label{defokforce}
 O_k = \frac{\Delta^\nu }{ \Delta R}
\log (J^{1/2}_{\mathbf{\Delta R}} \Psi_{\mathbf{\Delta R}} )
\end{equation}
The above  operators (\ref{defokforce}) are used  
also in the definition of the reduced matrix $\bar s$ 
for those elements depending on the variation with respect to a nuclear
coordinate. In this way it is possible to optimize both the wave function 
and the ionic positions at the same time, 
in  close analogy with the Car-Parrinello\cite{car} method applied to
the minimization problem. Also Tanaka \cite{tanaka} tried to perform 
Car-Parrinello like simulations via QMC, within the less efficient steepest
descent framework. 

An important  source of systematic errors is the dependence of the
variational parameters $c_i$ on the ionic configuration $\mathbf{R}$,
because for the final equilibrium geometry 
all the forces $f_i$ corresponding  to $c_i$ 
have to be zero, in order to guarantee that the true minimum 
of the potential energy surface (PES) is reached \cite{mella}.
As shown clearly in the previous subsection, within a QMC approach 
it is possible to control this condition by increasing systematically 
the bin length, when the thermal bias $T_{noise}$ vanishes. 
\begin{figure}[!ht]
\includegraphics[width=8.5cm]{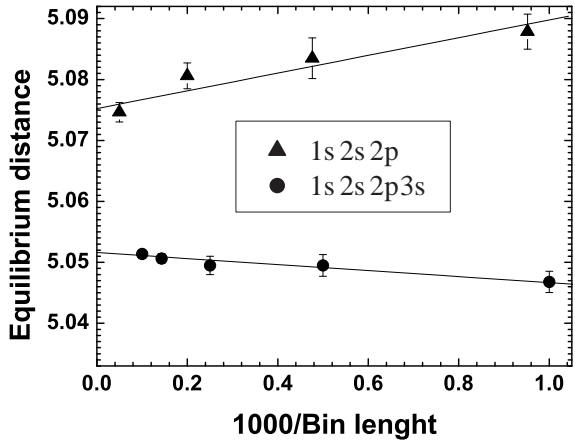}
\caption{\label{Li2}
Plot of the equilibrium distance of the $Li_2$ molecule as a function 
of the inverse bin length. The total energy and 
the binding energy are  reported in Tables \ref{benergies} and \ref{blengths}
respectively. 
 The triangles (full dots) refer to a simulation performed using $1000$
 ($3000$) iterations with $\Delta t= 0.015 H^-1$ ($\Delta t= 0.005 H^-1$)  and
 averaging over the last $750$ ($2250$) iterations. For all simulations the
 initial wavefunction is optimized  at $Li-Li$ distance $6$ a.u.}
\end{figure}
In Fig.~\ref{Li2} we report the equilibrium distance 
of the Li molecule as a function of the inverse bin length, 
so that an accurate evaluation of such an important quantity 
is possible even when the number of variational parameters is rather large, 
by extrapolating the value to an infinite bin length. 
However, as it is seen in the 
picture, though the inclusion of the 3s orbital in the atomic AGP basis 
substantially improves the equilibrium distance and the total energy by 
$\simeq 1mH$, this larger basis makes our simulation less efficient, 
as the time step $\Delta t$ has to be reduced by a factor three.  

We have not attempted to extend the geometry optimization to the 
more accurate DMC, since there are technical difficulties \cite{moroni}, 
and it is computationally much more demanding.

\subsection{Stochastic versus deterministic minimization}

In principle,
within a stochastic approach, the exact minimum is never reached 
as the forces $f_i$ are known only statistically with some error bar 
$\Delta f_i$. 
We have found that the method becomes efficient when all the forces 
are non zero only within few tenths of standard deviations. 
Then for small enough constant $\Delta t$, 
and large enough bin compatible with the computer resources, the 
stochastic minimization, obtained with a statistical evaluation of 
$\bar s$ continues in a stable manner, and 
 all the variational parameters fluctuate  after several iterations 
around a mean value.
After averaging these  variational parameters, the corresponding mean values 
 represent very good estimates satisfying the 
  minimum energy condition.  This can be verified by performing  
 an independent Monte Carlo simulation  much longer than the 
bin used for a single iteration of the stochastic  minimization, 
and then by explicitly checking that all the forces $f_i$ are zero within 
the statistical accuracy. 
 An example is given in Fig~\ref{forzap19} and 
discussed in the previous subsection.
 
On the other hand, whenever few  variational 
 parameters are clearly out of minimum, with $ |f_i /\Delta f_i|
 > \sigma_{cut}$, with $\sigma_{cut} \simeq 10$,
 we have found  a faster convergence with a much larger $\Delta t$, 
by applying the minimization 
scheme only for those selected parameters such that $ |f_i /\Delta f_i| >
\sigma_{cut}$, until $ |f_i /\Delta f_i| <\sigma_{cut}$.
After this initialization  it is then convenient  
to  proceed with the 
global minimization with all parameters changed at each iteration, 
in order to explore the variational space in a much more effective way. 

\subsection{ Different energy scales}
\label{multilevels}

 The SR method performs generally very well,
whenever   there is  only one energy scale in the 
variational wave function.  
However if  there are several energy scales in the problem, 
some of the variational parameters, e.g. the ones defining the 
low energy  valence orbitals, converge very slowly with respect to the
others, and the number 
of iterations required for the equilibration becomes exceedingly large,
considering also that the time step  $\Delta t$ necessary  for a stable 
convergence  depends 
on the high energy orbitals, whose dynamics cannot be accelerated beyond a
certain threshold. 
 It is easy to understand that SR technique not necessarily becomes inefficient for extensive systems with large number of atoms. Indeed suppose that we have $N$ atoms very far apart so that
we can neglect the interaction between electrons belonging to different atoms, than it is easy to see that the stochastic matrix Eq. \ref{defsr} factorizes in $N$ smaller matrices and the $\Delta t$ necessary for the convergence is equal to the single atom case, simply because the variational parameters of each single atom can evolve independently form each other. This is due to the size consistency of our trial function that can be factorized as a product of $N$ single atom trial functions in that limit. 
Anyway for system with a too large energy spread a way to overcome this difficulty was presented recently in 
Ref.~\onlinecite{filippi2}. 
Unfortunately this  method is limited to the 
optimization of the variational parameters  
in a super-CI-basis, to which   
a  Jastrow term is applied, that however   
can not be optimized together with the CI coefficients. 

In the present work, limited to a rather small 
 atomic basis, the SR technique is efficient, 
and general enough to perform the simultaneous 
optimization of the Jastrow and the determinantal part of the wave function, 
 a very important  feature 
that allows to capture the most non trivial correlations contained in 
our variational ansatz. 
Moreover, SR has been extended to deal with the structural optimization of a
 chemical system, which is another appealing characteristic of this method.
The results presented in the next section show that in some non trivial cases
the chemical accuracy can be reached also within this framework.

 However we feel that an improvement along the  line described in
 Ref.~\onlinecite{filippi2} 
will be  useful for realistic  electronic simulations  of complex systems 
with  many atoms, or when a very high precision is required at the
 variational level and consequently 
a wide spread of energy scales has to be included in
 the atomic basis.
We believe that the difficulty to work with a large 
basis set will be possibly alleviated  by using  pseudopotentials that allows to avoid the high energy 
components 
of the core electrons. However more work is necessary to clarify  the efficiency of the SR minimization scheme described here.

\section{Application  of the JAGP to molecules}

In this work we study total, 
correlation, and atomization energies, accompanied with the determination of
the ground state optimal structure for a restricted ensemble of molecules.
For each of them we performed a full all-electron SR geometry optimization,
starting from the experimental molecular structure. After the energy
minimization, we carried out all-electron VMC and DMC  simulations 
at the optimal geometry within the so called ''fixed node approximation''.  
The basis used here is a double zeta Slater set of atomic orbitals
(STO-DZ) for the AGP part, while for parameterizing the 3-body Jastrow geminal 
we used a double zeta Gaussian atomic set
(GTO-DZ). In this way both the antisymmetric and the bosonic part are well
described, preserving the right exponential behavior for the former and the
strong localization properties for the latter. 
Sometimes, in order to improve
the quality of the variational wave function we employed a mixed Gaussian and
Slater basis set in the Jastrow part, that allows to avoid a too strong dependency in the
variational parameters in a simple way. 
However, both in the AGP and in the
Jastrow sector we never used  a large basis set, in order to keep the 
wave function as simple as possible.  
The  accuracy of our wave function can be obviously improved by an extension of
the one particle  
basis set. but, as discussed in the previous section, this is rather difficult 
for a  stochastic minimization of the energy. Nevertheless,  
for most  of the molecules studied with a simple JAGP wave function, 
 a DMC calculation 
 is able to reach the chemical accuracy in the
binding energies and the SR optimization yields very precise geometries
already at the VMC level.

In the first part of this section some results will be presented for a small
set of widely studied molecules and belonging to the G1 database. 
In the second subsection we will treat the benzene and its radical cation $C_6
H_6^+$, by taking into account its distortion due to the Jahn-Teller
effect, that is  well reproduced by our SR geometry optimization.  

\subsection{Small diatomic molecules, methane, and water}

Except from $Be_2$ and $C_2$, all the molecules presented here belong to the
standard G1 
reference set; all their properties are well known and well reproduced by
standard quantum chemistry methods, therefore they constitute a good case for
testing new approaches and new wave functions.

The $Li$ dimer is one of the easiest molecules 
to be studied after the $H_2$, which is
exact for any Diffusion Monte Carlo (FN DMC) calculation with a
trial wave function that preserves the nodeless structure. $Li_2$ is less
trivial due to the presence of core electrons that are only partially
involved in the chemical bond and to the $2s-2p$ near degeneracy for the
valence electrons. Therefore many authors have done benchmark
calculation on this molecule to check the accuracy of the method or
to determine the variance of the inter-nuclear force calculated within a
QMC framework. In this work we start from $Li_2$ to move toward a structural
analysis of more complex compounds, thus showing  
that our QMC approach is able to handle relevant  chemical
problems. In the case of $Li_2$, a $3s~1p~STO-DZ$ AGP basis and a $1s~1p~GTO-DZ$
Jastrow basis turns out to be 
 enough for  the chemical accuracy (see the Appendix~\ref{lithium} for a
 detailed description of the trial wave function). More than $99\%$ 
of the correlation energy is recovered by a DMC simulation
(Table \ref{energies}), and the atomization
energy is exact within few thousandth of eV ($0.02~kcal~mol^{-1}$) (Table
\ref{benergies}).

Similar
accuracy have  been previously reached 
 within a DMC approach\cite{filippimol},  only by 
using a  multi-reference CI like wave function, 
that before our work,  was the usual 
way to improve the electronic nodal structure.
 As stressed before, the JAGP wave function includes many resonating
configurations through the geminal expansion, beyond the $1s~2s$ HF ground
state. 
 The bond length has been
calculated at the variational level through the fully optimized JAGP wave
function: the resulting equilibrium geometry 
turns out to be  highly accurate (Table \ref{blengths}), with
a discrepancy of only $0.001 a_0$ from the exact result.
 For this molecule it is worth  comparing our work 
with the one by Assaraf and Caffarel \cite{assaraf}. 
Their zero-variance zero-bias principle has been proved to be effective in
reducing the fluctuations related to the inter-nuclear force; however they
found that only the inclusion of the space warp transformation drastically
lowers the force statistical error, which magnitude becomes 
equal or even lower than the energy statistical error, thus allowing a feasible
molecular geometry optimization. 
Actually, our way of computing forces (see Eq.~\ref{vmcforce}) provides 
slightly  larger  variances,  without explicitly invoking the zero-variance
zero-bias principle.

The very good bond length, we obtained, is probably due to two main ingredients of our calculations:  we have carried out a stable energy optimization that is often more effective than the variance one, as shown by different authors \cite{varianza}, and we have used very accurate trial function as it is apparent from the good variational energy.


Indeed within our scheme we obtain good results without exploiting 
 the  computationally much more demanding DMC,  
thus highlighting the importance of  the SR minimization described in
Subsection~\ref{structural}.

Let us now consider larger molecules.
Both $C_2$ and $O_2$ are poorly described by a single Slater
determinant, since the presence of the nondynamic correlation is 
strong. 
Instead with a single geminal JAGP wave function, including implicitly  
  many Slater-determinants\cite{casula}, 
it is possible to obtain a quite good 
description of their molecular properties. For $C_2$, we 
used a $2s~1p~STO-DZ$ basis in the geminal, 
and a $2s~1p~DZ$ Gaussian Slater mixed basis in the Jastrow, for $O_2$ we
employed a $3s~1p~STO-DZ$ in the geminal and the same Jastrow basis as
before. In both
the cases, the variational energies recover more than $80\%$ of the correlation
energy, the DMC ones yield more than $90\%$, as shown in Tab.~\ref{energies}.  
These results are of the same level of accuracy as  
those obtained by Filippi \emph{et al}\cite{filippimol} with a multireference 
wave function by using the same Slater basis for the antisymmetric part 
and a different Jastrow factor. 
>From the Table \ref{benergies} of the atomization
energies, it is apparent that DMC considerably improves the binding
energy with respect to the VMC values, 
although for these two molecules it is quite far from the chemical
accuracy ($\simeq$ 0.1 eV): for $C_2$ the error is 0.55(3) eV, for $O_2$
it amounts to 0.67(5) eV. Indeed, it is well known that  the electronic 
structure of the atoms is described better than the corresponding molecules if
the basis set remains the same, and the nodal error is not compensated
by the energy difference between the separated atoms and the molecule.
In a benchmark DMC calculation with pseudopotentials \cite{grossman}, 
Grossman found an error
of 0.27 eV in the atomization energy for $O_2$, by using a single determinant
wave function; probably, pseudopotentials allow the error 
between the pseudoatoms and the pseudomolecule to compensate better, 
thus yielding more accurate energy differences.
As a final remark on the $O_2$ and $C_2$ molecules, 
our bond lengths are in 
between the LDA and GGA precision, and 
still worse than the best CCSD calculations,
but  our results may be considerably improved by a larger atomic basis 
set, that we have not attempted so far. 

Methane and water are very well described by the JAGP wave function. Also for
these molecules we recover more than $80\%$ of correlation
energy at the VMC level, while DMC yields more than $90\%$, 
with the same level of accuracy reached in previous Monte Carlo studies
\cite{huang,garmer,lu,luchow}. Here the binding energy
is almost exact, since in this case the nodal energy error arises essentially
from only one atom (carbon or oxygen) 
and therefore it is exactly compensated when the
atomization energy is calculated. Also the bond lengths are highly accurate,
with an error lower then 0.005 $a_0$.

For $Be_2$ we applied a 3s 1p STO-DZ basis set for the AGP part and a 2s 2p DZ
Gaussian Slater mixed basis for the Jastrow factor. VMC calculations 
performed with this
trial function at the experimental equilibrium geometry yield $90\%$ of the
total correlation energy, while DMC gives
$97.5\%$ of correlation, i.e. a total energy of -29.33341(25) H. 
Although this value is
better than that obtained by Filippi \emph{et al} \cite{filippimol} (-29.3301(2)
H) with a smaller basis ($3 s$ atomic orbitals not included), it is not enough
to bind the molecule, because the binding energy remains still positive
(0.0069(37) H). Instead, once the molecular geometry has been relaxed,
the SR optimization finds a bond distance of $13.5(5)~a_0$ 
 at the VMC level;
therefore the employed basis allows the molecule to have a Van der Waals like
minimum, quite far from the experimental value.
In order to have a reasonable description of the bond length and the atomization
energy, one needs to include at least a $3s 2p$ basis in the antisymmetric
part, as pointed out in Ref.~\onlinecite{martin}, and indeed  an atomization
energy compatible with the experimental result (0.11(1) eV) has been obtained
within the extended geminal model \cite{roeggen} by using a much larger basis
set (9s,7p,4d,2f,1g) \cite{roeggen2}. 
This  suggests that a complete basis set  calculation with  JAGP 
may describe also this  molecule.  
 However, as already mentioned in subsection
\ref{multilevels}, our SR method can not cope with  a very large basis in a
feasible computational time. Therefore we believe that at present 
the accuracy needed to describe correctly $Be_2$ is  
out of the possibilities of the approach. 

\subsection{Benzene and its radical cation}

\begin{figure}[!ht]
\includegraphics[width=9cm]{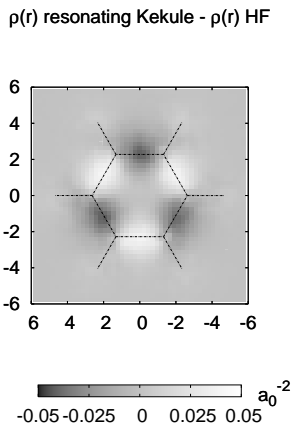}
\caption{\label{firHFdiffden}
 Surface plot of the charge 
density projected onto the molecular plane. 
 The difference between the non-resonating
(indicated as HF) and resonating Kekul\'e 3-body Jastrow wave function 
densities is shown.
Notice the corresponding change  
from a dimerized structure to a $C_6$ rotational invariant density
profile.}
\end{figure}

We studied the $^1 A_{1g}$ ground state of the benzene molecule by using a
very simple one particle basis set: for the AGP, a 2s1p DZ set centered 
on the carbon atoms and a 1s SZ on the hydrogen, 
instead for the 3-body Jastrow, a 1s1p DZ-GTO  set centered only on the carbon
sites. $C_6 H_6$ is a peculiar molecule, since its highly symmetric ground
state, which belongs to the $D_{6h}$ point group, is a resonance among
different many-body states,  each of them characterized by 
 three double bonds between carbon atoms. 
This resonance is responsible for the stability of the
structure and therefore for its aromatic properties. We started from a non
resonating 2-body Jastrow wave function, which dimerizes the ring and breaks
the full rotational symmetry, leading to  the Kekul\'e configuration.  
As we expected, the inclusion of the resonance between the two possible
Kekul\'e states lowers the VMC energy by more than 2 eV. The wave function
is further improved by adding another type of resonance, that includes also the
Dewar contributions connecting third nearest neighbor carbons.
 As reported in Tab.~\ref{benzene}, the
gain with respect to the simplest Kekul\'e  wave function amounts to 4.2 eV, 
but the main improvement arises from the further 
inclusion of the three body Jastrow
factor, which allows to recover the $89 \%$ of the total atomization energy at
the VMC level. 
The main effect of the three body term is to 
keep  the total charge around the carbon sites  to approximately six electrons, 
thus penalizing   the double  occupation of the 
 $p_z$ orbitals. The same important correlation ingredient is present  
in the well known  Gutzwiller wave function already used for polyacetylene
\cite{gutz}. 
Within this scheme we  have systematically included 
in the 3-body Jastrow part 
  the same type of terms present in  the AGP one,   
namely both  $g^{a,b}$    
and  $\lambda^{a,b}$ are non zero for the same pairs  of atoms. 
 As expected, the  terms connecting  next nearest neighbor  carbon 
sites are much less 
important than the remaining ones
 because the VMC energy  does not significantly improve
(see  the full resonating + 3-body wave function in Tab.~\ref{benzene}).
A more clear  behavior is found by
carrying out DMC simulations: the interplay between the resonance among
different structures and the Gutzwiller-like correlation 
refines more and more the nodal surface topology, 
thus lowering the DMC energy by significant amounts. 
Therefore it is crucial to insert
into the variational wave function all these ingredients in order to have an
adequate description of the molecule. For instance, in
Fig. \ref{firHFdiffden} we report the density surface difference between the 
non-resonating 3-body Jastrow wave function, which breaks 
the $C_6$ rotational invariance, and the resonating Kekul\'e
structure, which preserves the correct $A_{1g}$ symmetry: the change in the
electronic structure is significant.
The best result for the binding energy  is obtained with the
Kekul\'e Dewar resonating 3 body wave function, which recovers the $98,6\%$ of
the total atomization energy with an absolute error of 0.84(8) eV.
As Pauling \cite{pauling} first pointed out, benzene is a genuine RVB system,
indeed it is well described by the JAGP wave function.
 Moreover Pauling gave an estimate for the resonance energy of 1.605 eV from
 thermochemical experiments in qualitative agreement with our results.
A final  remark about the error in the total atomization energy: the latest
frozen core CCSD(T) calculations \cite{feller2,srinivasan} are able to reach a
precision of 0.1 eV, but only after the complete basis set extrapolation and the
inclusion of the core valence effects to go beyond the pseudopotential
approximation.  Without the latter  corrections, the error is quite
large even in the CCSD approach, amounting to 0.65 eV \cite{feller2}. In our
case, such an error arises from the fixed node approximation, whose nodal
error is not compensated by the difference between the atomic and the
molecular energies, as already noticed in the previous subsection.

\begin{figure}[!ht]
\includegraphics[width=8.5cm]{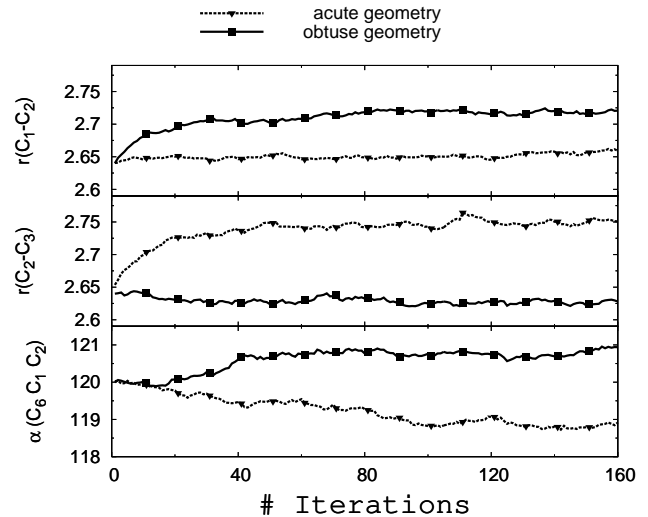}
\caption{\label{relax}
Plot of the convergence toward the equilibrium geometry for the $^2 B_{2g}$
acute and the $^2 B_{3g}$ obtuse benzene cation. Notice that both the
simulations start form the ground state neutral benzene geometry and relax
with a change both in the $C-C$ bond lengths and in the angles. The symbols 
are the same of Tab.~\protect\ref{benzenelengths}.}
\end{figure}

The radical cation $C_6 H_6 ^+$ of the benzene molecule has been the subject
of intense theoretical studies\cite{klaus,deleuze}, aimed to focus on the 
Jahn-Teller distorted ground state structure. 
Indeed the ionized $^2 E_{1g}$ state, which
is degenerate, breaks the symmetry and experiences a relaxation from the
$D_{6h}$ point group to two different states, $^2 B_{2g}$ and $^2 B_{3g}$,
that belong to the lower $D_{2h}$ point group. In practice, the former is the
elongated acute deformation of the benzene hexagon, the latter is its
compressed obtuse distortion. We applied the SR
structural optimization, starting from the $^2 E_{1g}$ state, and the
minimization correctly yielded a deformation toward the
acute structure for the $^2 B_{2g}$ state and the obtuse for the $^2 B_{3g}$
one; the first part of the 
evolution of the distances and the angles during those simulations is
shown  in Fig.\ref{relax}. 
After this equilibration, average over 200 further 
iterations yields  bond
distances and angles with the same accuracy as the all-electron 
BLYP/6-31G* calculations reported in Ref.~\onlinecite{klaus} (see Tab.~
\ref{benzenelengths}).
 As it appears from Tab.~\ref{ionized} 
 not
only the structure but also the DMC total energy is in perfect agreement
with the BLYP/6-31G*, and much better than 
SVWN/6-31G* that does not contain semi empirical functionals, for which the 
comparison with our calculation is more appropriate, being fully ab-initio. 

The difference of the  VMC and DMC energies between 
 the two distorted cations are the same within the
error bars; indeed, the determination of which structure is the real cation
ground state is a challenging problem, since the experimental results give a
difference of only few meV in favor of the obtuse state and also the most
refined quantum chemistry methods are not in agreement among themselves
\cite{klaus}. A more affordable problem is the determination of the
adiabatic ionization potential (AIP), calculated for the $^2 B_{3g}$ state,
following the experimental hint. Recently, very precise CCSD(T) calculations
have been performed in order to establish a benchmark theoretical study for the
ionization threshold of benzene \cite{deleuze}; the results are reported in
Tab.~\ref{adiabatic}. After the correction of the zero point
energy due to the different structure of the cation with respect to the
neutral molecule and taken from a B3LYP/cc-pVTZ calculation reported in
Ref.~\onlinecite{deleuze}, the agreement among our DMC result, the benchmark
calculation and the experimental value is impressive. Notice that in this case  
there should be a perfect cancellation of nodal errors in order to obtain 
 such an accurate value; however,  we believe that 
it is not a fortuitous result, because in this case the
underlying nodal structure does not change much by  adding or removing a single 
electron.
 Therefore we expect that this property holds for
all the affinity and ionization  
 energy calculations with  a particularly accurate 
variational wave function as the one we have proposed  here.  
Nevertheless  DMC is needed to reach the chemical accuracy, since
the VMC result is slightly off from the experimental one just by few 
tenths of eV. The AIP and the geometry determination for the $C_6 H_6 ^+$ are
encouraging to pursue this approach, with the aim to describe 
even much more interesting and challenging   chemical systems.

\section{Conclusion}
In this work, we have tested the JAGP wave function on simple molecular 
systems where  accurate results are known.
As shown in the previous section  
a large amount of the correlation energy  is already  recovered 
at the variational level with a  computationally very 
efficient and feasible method, 
extended in this work to the nuclear 
geometry optimization. Indeed, 
 much larger systems  should be  
tractable because, within the JAGP ansatz,  it is 
sufficient  to sample a single determinant 
whose dimension scales only with the number of electrons. 
The presence of the  Jastrow factor implies the evaluation of  
multi-dimensional integrals  that,  so far
can be calculated efficiently  only with the Monte Carlo method. 
Within this framework, 
it is difficult to reach  the complete  basis set limit,
both in the Jastrow and the AGP terms, 
although some  progress has been made recently\cite{fahy2,filippi2}. 
Even if the dimension of the basis is  limited by the 
difficulty to perform energy optimization with a very large number of 
variational parameters, 
we have obtained the chemical accuracy for most cases studied. 
From a general point of view  
the basis set convergence of the JAGP is expected to be faster than AGP 
considering that  the electron-electron cusp condition 
 is  fulfilled exactly at each level  of the expansion.  
Nevertheless, all results presented here can be systematically improved 
with larger basis set.
In particular
the $Be_2$ bonding distance  should be substantially   corrected  by   
a more complete basis, that we have not attempted so far.\cite{roeggen2}

The usefulness of the JAGP wave function is already well known in 
the study of strongly correlated systems defined on a lattice.
For instance in the widely studied  Hubbard model, as well as  in any model 
with electronic repulsion, it  is not possible to obtain 
a superconducting ground state at the mean-field Hartree-Fock level.
On the contrary, as soon as a correlated Jastrow term is 
applied to the BCS wave function (equivalent to the AGP wave function
in momentum space\cite{bcsagp}),
 the stabilization of a d-wave superconducting order 
parameter is possible, and is expected to be a realistic property of 
the model\cite{hubbard}. 
More interestingly the presence of the Jastrow factor can qualitatively  change 
the wave function especially at one electron per site filling, 
by converting  a BCS superconductor to a Mott insulator with 
a finite charge gap\cite{capello}.

The same effect is 
clearly seen for the gedanken experiment  of a dilute gas of $H_2$ molecules, 
a clarifying test example already  used  in the introduction. 
The AGP wave function is essentially exact for a single molecule (at least 
with the complete basis set), but its obvious size consistent 
 extension to the gas 
would lead to the unphysical result of superconductivity because the 
charge around each molecule  would be  free to fluctuate within the 
chosen set of geminal orbitals. Only the presence of the Jastrow term 
added to this wave function, 
allows the local conservation of the charge around the molecule, by  forbidding 
unphysical $H_2$ dimers  with more than two electrons. 
Once the charge is locally conserved, the phase of the BCS-AGP 
 wave function cannot 
have a definite value and phase coherence is correctly forbidden  by 
the Jastrow factor.
In the present work,  the interplay between the Jastrow and the geminal part
has been shown to be very effective  
in all cases studied and particularly   
  in the non trivial case of the benzene molecule, where we have 
shown systematically the various approximations.
Only when {\em both}  the Jastrow and the AGP  terms 
are accurately optimized together, 
the  AGP nodal structure of the wave function is considerably improved. 
For the above reasons and the size consistency of the JAGP  we 
expect that this wave function 
should be  generally accurate also in complex systems 
made by many molecules. 
 The local conservation of the charge around each molecule  
is taken into account 
 by the Jastrow factor, whereas the quality  of each molecule  is 
described  also by the AGP geminal part exactly as in the $H_2$ gas example.

In the near future it is very appealing and promising 
to extend the  JAGP  study to the  DNA nitrogenous bases, 
whose geometrical structure is very similar to the benzene ring.
In particular  we plan to accurately evaluate 
the energetics (reduction potential, ionization energies, electron 
affinity, etc.) of DNA bases  and base pairs, quantities of great importance 
to characterize excess electron and hole transfer which are involved in 
radiation damage as well as in the development of novel DNA technologies.

\appendix

\section{ Stabilization of the SR technique}
\label{srstable}
Whenever the number of variational parameters increases, it often happens 
that the stochastic $(N+1)\times (N+1)$ matrix
\begin{equation} \label{defsrapp}
s_{k,k^\prime} ={ 
 \langle  \Psi | O_k O_{k^\prime} | \Psi\rangle \over 
\langle \Psi | \Psi \rangle }
\end{equation}
becomes singular, i.e. the condition number, defined as the ratio
$\sigma=\lambda_N/\lambda_1$ between its maximum $\lambda_N$  and  
minimum eigenvalue $\lambda_1$,  is too large. In that case 
the inversion of this matrix generates clear numerical instabilities 
which are difficult to control especially within a statistical method.
Here $O_k={  d {\rm ln} \Psi (x) \over d \alpha_k } $ 
are the operators corresponding to the variational parameters 
$\alpha_k$ appearing in the wave function $\Psi$ 
 for $k=1,\cdots N$, whereas 
for $k=0$ the operator $O_0$ represents the identity one. 

The first successful proposal to control this instability was to remove from 
the inversion problem (\ref{srstandard}), required for the minimization, those  
 directions in the variational parameter space corresponding to exceedingly 
small eigenvalues $\lambda_i$.   

 In this appendix we describe a better method.
 As a first step,  we show that the reason of the
large condition number $\sigma$  is due to the existence of ''redundant'' 
variational parameters that do not make changes to the wave function 
within a prescribed tolerance $\epsilon$. 
Indeed in practical calculations, we are interested in the minimization 
of the wave function within a reasonable  accuracy. 
The tolerance $\epsilon$ may represent therefore the distance
 between the exact normalized variational wave function  which 
minimizes the energy expectation value  and the  
approximate acceptable one, 
for which we no longer iterate the minimization 
scheme. For instance  $\epsilon=1/1000$ is by far acceptable for  
 chemical and physical interest. 
A stable algorithm is then obtained by simply 
removing the parameters that do not change the wave function 
by less than $\epsilon$  from the minimization.  
An efficient scheme to remove the ''redundant parameters'' is also given.

Let us consider the $N$ normalized states orthogonal 
to $\Psi$, but not orthogonal among each other:
\begin{equation} \label{vectors}
|e_i\rangle = { (O_k - s_{k,0}) | \Psi \rangle  \over 
 \sqrt { \langle \Psi |  (O_k - s_{k,0})^2 |\Psi } }. 
\end{equation}
where $s_{k,0}$ is defined in Eq.~\ref{defsrapp}.
These  normalized vectors define $N$ directions
in the $N-$dimensional  variational parameter manifold, 
which are independent 
as long as the determinant $S$ of the corresponding $N\times N$ overlap matrix 
\begin{equation}
\label{overlap}
\bar s_{k,k^\prime}  = \langle e_k | e_{k^\prime} \rangle 
\end{equation}
is non zero. The number $S$ is clearly  positive 
and it assumes its maximum value $1$   whenever  
all the directions $e_i$  are mutually orthogonal. 
On the other hand, let us suppose 
that there exists an eigenvalue $\bar \lambda$  of 
$\bar s$  smaller than the square of the desired 
tolerance $\epsilon^2$,
then the corresponding eigenvector  $|v>=\sum_i a_i  |e_i\rangle$ 
 is such that:
\be
\langle v | v \rangle =  \sum_{i,j} a_i a_j \bar s_{i,j} 
= \bar \lambda
\ee 
where the latter equation holds due to the normalization 
condition $\sum_i a_i^2 =1$. 
We arrive therefore to the
conclusion that it is possible to define a vector $v$  with almost  
vanishing norm $|v| =\sqrt{\lambda} \le \epsilon$ as a linear combination of   
 $e_i$, with at least some non zero coefficient. 
This implies that the $N$  directions $e_k$  
are linearly dependent within a tolerance $\epsilon$ and one can safely 
remove at least one  parameter from the calculation.

In general whenever there 
 are $p$ vectors $v_i$ that are below  the tolerance $\epsilon$ 
the optimal choice to stabilize the minimization procedure is 
to remove $p$ rows and $p$ columns from the matrix (\ref{overlap}), 
in such a way that the corresponding determinant of the $(N-p) \times (N-p)$ 
overlap matrix  is maximum.

>From practical purposes it is enough to consider an iterative scheme 
to find a large minor, but  not necessarily the maximum one. 
This method  is based on the inverse of $\bar s$. At each step we remove the 
$i-th$ row and column from $\bar s$  for which $\bar s^{-1}_{i,i}$ is 
maximum. We stop to remove rows and columns after $p$ inversions.
In this approach we exploit the fact that, by a consequence of
the Laplace theorem on determinants,  
$\bar s^{-1}_{k,k}$ is the ratio between the described minor without 
the $k-th$ row and column and the determinant of the full 
$\bar s$ matrix.
Since within a stochastic method it is certainly not possible to work with a 
machine precision tolerance, setting  
 $\epsilon=0.001$  guarantees a stable algorithm, without affecting 
the accuracy of the calculation. 
 The advantage of this scheme, compared with  the previous one\cite{sorella}, 
is that the less relevant parameters can be easily  identified
after few iterations and do not change further in the process of minimization.

\section{Size consistency of the 3-body Jastrow factor}
\label{consistency}
In order to prove the size consistency property of the three body Jastrow factor
described in Sec.~\ref{3bodysec}, let us take into account 
a system composed by two well separated subsystems $A$ and
$B$, which are distinguishable and whose dimensions are much smaller than the
distance between themselves; in general they may contain more then one atom.
In this case the Jastrow function $J_3$ (\ref{3body})  can be
written as $J_3=e^{U}$ with:
\begin{eqnarray}
\label{sum}
U & = & \frac{1}{2} \sum_{\begin{array}{c} 
i,j \in A \\i \ne j
\end{array}} \phi(r_i,r_j) +
\frac{1}{2} \sum_{\begin{array}{c} 
i,j \in B \\
i \ne j
\end{array}} \phi(r_i,r_j) \\ \nonumber 
 & + &
\sum_{i \in A} \sum_{j \in B} \phi(r_i,r_j),
\end{eqnarray}
where we have explicitly considered the sum over different subsystems.
As usual, the two particle function $\phi(r_i,r_j)$ is 
expanded over a single particle basis $\psi$, centered on each nucleus of
the system: 
\begin{equation}
\label{gem}
\phi(r_i,r_j)=\sum_{m,n} \lambda^{m,n} \psi^m(r_i) \psi^n(r_j).
\end{equation}
The indices $n$ and $m$ refer not only to the basis elements but also 
to the nuclei which the orbitals are centered on.

The self consistency problem arises from the last term in Eq.~\ref{sum},
i.e. when the electron $r_i$ belongs to $A$ and $r_j$ to $B$. If the Jastrow
is size consistent, whenever $A$ and $B$ are
far apart from each other this term must vanish or at 
most generate a one body term that is in turn size consistent, as 
we are going to show in the following.
In the  limit of large separation 
all  the $\lambda^{m,n}$ off diagonal terms connecting any
basis element of $A$ 
to any basis element of $B$ must   vanish. 
The second requirement is  a sufficiently fast 
 decay of the basis set orbitals
$\psi(r)$ whenever $r \rightarrow \infty$,
except at most 
for a constant term $C_n$ which may be present in the single particle
orbitals, and  is useful  to improve the variational energy. 

For the sake of generality, suppose that the system $A$
contains $M_A$ nuclei and $N_A$ electrons. The first requirement implies that:
\begin{eqnarray}
\phi(r_i,r_j) & = & \sum_{m,n \in A} \lambda^{m,n} \psi^m(r_i) \psi^n(r_j) 
\\ \nonumber
& + &
\sum_{m,n \in B} \lambda^{m,n} \psi^m(r_i) \psi^n(r_j),
\end{eqnarray}
instead the second allows to write the following expression for the mixed term
in Eq.~\ref{sum}: 
\begin{equation}
\label{mixed}
\sum_{i \in A} \sum_{j \in B} \phi(r_i,r_j) = N_B \sum_{n \in A} C_n P_n +
N_A \sum_{m \in B} C_m P_m,
\end{equation}
where the factors $P_n$ are one body terms defined as:
\begin{equation}
P_n= \left \{  \begin{array}{l}
\sum_{m \in A} \lambda^{n,m} \sum_{i \in A} \psi^m(r_i) \textrm{~~~if $n \in A$}
\\
\sum_{m \in B} \lambda^{n,m} \sum_{i \in B} \psi^m(r_i) \textrm{~~~if $n \in B$}
\end{array} \right .
\end{equation}
Notice that if all the orbitals decay to zero,
 the size consistency is immediately recovered, since
the sum in Eq.~\ref{mixed} vanishes. 
Analogously to the derivation we have done to extract 
the one body contribution from the mixed term, 
the other two terms on the RHS of Eq.~\ref{sum}
can be rearranged in the following form: 
\begin{eqnarray}
\frac{1}{2} \sum_{\begin{array}{c} 
i,j \in A \\
i \ne j
\end{array}} \phi(r_i,r_j) & =  & (N_A-1) \sum_{n \in A} C_n P_n
\\ \nonumber
 & + & \textrm{~~two body terms}, 
\end{eqnarray}
and the sum in Eq.~\ref{sum} can be rewritten as:
\begin{eqnarray}
U & = & (N-1)\sum_{n \in A} C_n P_n+(N-1)\sum_{n \in B} C_n P_n
\\ \nonumber
& + & \textrm{~~two body size consistent terms}. 
\end{eqnarray}
Therefore the size consistency implies that
the scaling of the $C_n$ with the total number of particle $N$ is:
\begin{equation}
C_n = \frac{c_n}{N-1},
\end{equation}
as mentioned in Sec.~\ref{3bodysec}.

\section{An example case: JAGP wave function for $Li_2$}
\label{lithium}
We briefly describe the application of the JAGP to the $Li_2$ molecule. 
This example shows the beauty of our approach that allows to describe the
chemical bond as resonance of many pairing functions whose importance is
weighted by the $\lambda$ coefficients. 
In the expansion of the geminal function for the determinant in
Eq. \ref{expgem} we used six orbitals for each atom: 
\bea
\phi_{1s},\phi_{2s} &=& C_{1s} \left ( e^{-z_1r} + p e^{-z_2r} \right ), \\
 \phi_{2p} &=& C_{2p} \, \vec{r} e^{-z_1r}, \\ 
\phi_{3s} &=& C_{3s} \, r^2 e^{-z_1r}.
\eea
The parameters $p$ in $1s,2s$ orbitals are fixed by the single atomic cusp
conditions, and $C_{1s}, C_{2p}, C_{3s}$ are the normalization
constants. These orbitals are connected by different $\lambda_{m,n}^{a,b}$,
which obey the constraints given by the symmetry of the system, and are
reported in table \ref{lambda1}. Since the trial function does not need to be
normalized we  set $\lambda_{1s,1s}^{a,a}$  equal to $1$. The total number of
non zero $\lambda$ is 58, but the constraints allow to reduce them to 18
variational parameters.
\\
In the Jastrow part we used a two body term that is a slightly modified
version of the Eq. \ref{twobodyold}.
In fact due to the simple symmetry of the
system is possible to build a Jastrow more suitable for this 
diatomic molecule, namely: 
\be
u(r,z)=  \frac{r}{2 (1 +  \sqrt{a (x^2+y^2)+bz^2} ) },
\ee
which distinguishes the different components of the two electrons
distance. We found that this two body Jastrow factor is particularly useful
for $Li_2$, which is much more elongated than the other molecules studied
here, for which the usual form in Eq. \ref{twobodyold} has been employed. 
The optimal parameters
obtained  for the Jastrow are $a=0.8796\mbox{ , } b=0.7600$.  
In the expansion of the pairing function for the three body Jastrow term (see
Eq. \ref{3body}) we used the following orbitals: 
\bea
\phi_{sG} &=& e^{-z_1r^2} + p, \\ 
\phi_{pG} &=& \vec{r} \left ( e^{-z_1r^2} + p  e^{-z_2 r^2} \right ).
\eea
The $\lambda$ matrix that connects these orbitals is given in table
\ref{lambda2}; this matrix fulfills the same symmetry constraints as in the
case of the paring determinant. In this case the total number of non zero
$\lambda$ is 24 and the symmetry reduces the variational freedom to only 8
parameters.
The single particle orbitals are reported in table
\ref{parli2}, and include other 15 parameters.

\acknowledgments 
We grateful acknowledge P. Carloni, F. Becca and L. Guidoni 
 for  fruitful  suggestions in writing this manuscript. 
 We also thank 
 S. Moroni, S. Zhang,  C. Filippi and D. Ceperley for useful discussions. 
This work was partially supported by  MIUR, COFIN 2003.

\newpage

\begin{table*}[!hbp]
\caption{\label{energies}
Total energies in variational ($E_{VMC}$) and 
diffusion ($E_{DMC}$) Monte Carlo calculations; 
the percentages of correlation energy recovered in
VMC ($E^{VMC}_c(\%)$) and DMC ($E^{DMC}_c(\%)$) have been evaluated using
the ``exact'' ($E_0$) and Hartree--Fock ($E_{HF}$) energies from
the references reported in the table. Here ``exact'' means the ground state
energy of the non relativistic infinite nuclear mass hamiltonian.
The energies are in \emph{Hartree}.}
\begin{ruledtabular}
\begin{tabular}{l d d d d d d }
& \makebox[0pt][c]{$E_0$} & \makebox[0pt][c]{$E_{HF}$} &
\makebox[0pt][c]{$E_{VMC}$} & \makebox[0pt][c]{$E^{VMC}_c(\%)$} 
& \makebox[0pt][c]{$E_{DMC}$}  & \makebox[0pt][c]{$E^{DMC}_c(\%)$} \\
\hline
$Li$ & -7.47806 \footnotemark[1] & -7.432727 \footnotemark[1]
& -7.47721(11) &  98.12(24)  & -7.47791(12) &  99.67(27)  \\
$Li_2$ & -14.9954 \footnotemark[3] & -14.87152 \footnotemark[3] 
& -14.99002(12) &  95.7(1)  & -14.99472(17) &  99.45(14) \\
$Be$ & -14.66736 \footnotemark[1] & -14.573023 \footnotemark[1] 
& -14.66328(19) &  95.67(20) &  -14.66705(12) &  99.67(13) \\
$Be_2$ & -29.33854(5) \footnotemark[3] & -29.13242 \footnotemark[3] 
&  -29.3179(5) &  89.99(24)  &  -29.33341(25) & 97.51(12)   \\
$O$ &  -75.0673 \footnotemark[1]  & -74.809398 \footnotemark[1] 
& -75.0237(5) & 83.09(19) & -75.0522(3) & 94.14(11) \\
$H_2O$ & -76.438(3)  \footnotemark[2] & -76.068(1) \footnotemark[2]
& -76.3803(4) & 84.40(10) & -76.4175(4) & 94.46(10) \\
$O_2$ & -150.3268 \footnotemark[3] & -149.6659 \footnotemark[3] 
& -150.1992(5) & 80.69(7) & -150.272(2) & 91.7(3) \\
$C$ & -37.8450 \footnotemark[1] & -37.688619 \footnotemark[1]
& -37.81303(17) & 79.55(11)  & -37.8350(6) &  93.6(4)  \\
$C_2$ & -75.923(5) \footnotemark[3] & -75.40620 \footnotemark[3] 
& -75.8273(4) & 81.48(8)  & -75.8826(7)  & 92.18(14)  \\
$CH_4$ & -40.515 \footnotemark[4] & -40.219 \footnotemark[4] 
& -40.4627(3) & 82.33(10) & -40.5041(8) & 96.3(3)  \\
$C_6H_6$ & -232.247(4) \footnotemark[5] & -230.82(2) \footnotemark[6] 
& -231.8084(15) & 69.25(10)   & -232.156(3) &  93.60(21)  \\
\end{tabular}

\footnotetext[1]{Exact and HF energies from Ref.~\onlinecite{exact}.}
\footnotetext[2]{Ref.~\onlinecite{feller}.}
\footnotetext[3]{Ref.~\onlinecite{filippimol}.}
\footnotetext[4]{Ref.~\onlinecite{huang}.}
\footnotetext[5]{Estimated ``exact'' energy from Ref.~\onlinecite{srinivasan}.}
\footnotetext[6]{Ref.~\onlinecite{ermler}.}

\end{ruledtabular}
\end{table*}

\begin{table*}[!hbp]
\caption{\label{benergies}
Binding energies in $eV$ obtained by variational ($\Delta_{VMC}$) and diffusion
($\Delta_{DMC}$) Monte Carlo calculations; 
$\Delta_0$ is the ``exact'' result for the non-relativistic infinite nuclear
mass hamiltonian. 
Also the percentages ($\Delta_{VMC}(\%)$ and $\Delta_{DMC}(\%)$) 
of the total binding energies are reported.}
\begin{ruledtabular}
\begin{tabular}{l d d d d d }
& \makebox[0pt][c]{$\Delta_0$} & \makebox[0pt][c]{$\Delta_{VMC}$} &
\makebox[0pt][c]{$\Delta_{VMC}(\%)$} & \makebox[0pt][c]{$\Delta_{DMC}$} &
\makebox[0pt][c]{$\Delta_{DMC}(\%)$} \\
\hline
$Li_2$ & -1.069 & -0.967(3) & 90.4(3) & -1.058(5) & 99.0(5)  \\
$O_2$ & -5.230 & -4.13(4) & 78.9(8) & -4.56(5) &  87.1(9) \\
$H_2O$ & -10.087 & -9.704(24) & 96.2(1.0) & -9.940(19) & 98.5(9) \\
$C_2$ & -6.340 & -5.476(11) & 86.37(17) & -5.79(2) &  91.3(3) \\
$CH_4$ & -18.232 & -17.678(9) & 96.96(5)  & -18.21(4)  & 99.86(22)  \\
$C_6H_6$ & -59.25 & -52.53(4) & 88.67(7)  & -58.41(8) & 98.60(13) \\
\end{tabular}
\end{ruledtabular}
\end{table*}

\begin{table*}[!hbp]
\caption{\label{blengths}
Bond lengths ($R$) in atomic units; 
the subscript $0$ refers to the ``exact'' results.
For the water molecule $R$ is the distance between O and H
and $\theta$ is the angle HOH (in deg), for $CH_4$ $R$ is the distance between
C and H and $\theta$ is the HCH angle. }

\begin{ruledtabular}
\begin{tabular}{l d d d d }
& \makebox[0pt][c]{$R_0$} & \makebox[0pt][c]{$R$}  
& \makebox[0pt][c]{$\theta_0$} & \makebox[0pt][c]{$\theta$} \\
\hline
$Li_2$ &  5.051 &  5.0516(2) & & \\
$O_2$ & 2.282  &  2.3425(18)  &   & \\
$C_2$ & 2.348  & 2.366(2)  &   & \\
$H_2O$ &  1.809 &  1.8071(23) & 104.52 & 104.74(17) \\ 
$CH_4$ & 2.041 & 2.049(1) & 109.47 & 109.55(6) \\
\hline
& \makebox[0pt][c]{$R^{CC}_0$} & \makebox[0pt][c]{$R^{CC}$}  
& \makebox[0pt][c]{$R^{CH}_0$} & \makebox[0pt][c]{$R^{CH}$} \\
\hline
$C_6H_6$ &  2.640 & 2.662(4) & 2.028 & 1.992(2) \\
\end{tabular}
\end{ruledtabular}
\end{table*}

\begin{table*}[!hbp]
\caption{\label{benzene}
Binding energies in $eV$ obtained by variational ($\Delta_{VMC}$) and diffusion
($\Delta_{DMC}$) Monte Carlo calculations with different trial wave functions
for benzene. In order to calculate the binding energies yielded 
by the 2-body Jastrow we used the atomic energies reported 
in Ref.~\onlinecite{casula}.
The percentages ($\Delta_{VMC}(\%)$ and $\Delta_{DMC}(\%)$) 
of the total binding energies are also reported. }
\begin{ruledtabular}
\begin{tabular}{l d d d d }
& \makebox[0pt][c]{$\Delta_{VMC}$} &
\makebox[0pt][c]{$\Delta_{VMC}(\%)$} & \makebox[0pt][c]{$\Delta_{DMC}$} &
\makebox[0pt][c]{$\Delta_{DMC}(\%)$} \\
\hline
Kekul\'e + 2body & -30.57(5) & 51.60(8)  & - & -  \\
resonating Kekul\'e + 2body & -32.78(5) & 55.33(8) & - & - \\
resonating Dewar Kekul\'e + 2body & -34.75(5) & 58.66(8) & -56.84(11) &
95.95(18) \\ 
Kekul\'e + 3body & -49.20(4) & 83.05(7) & -55.54(10) &  93.75(17) \\
resonating Kekul\'e + 3body & -51.33(4) & 86.65(7) & -57.25(9)  & 96.64(15)  \\
resonating Dewar Kekul\'e + 3body & -52.53(4) & 88.67(7)  &  -58.41(8) &
98.60(13) \\ 
full resonating + 3body & -52.65(4) & 88.87(7) & -58.30(8) &  
98.40(13) \\

\end{tabular}
\end{ruledtabular}
\end{table*}

\begin{table*}[!hbp]\caption{\label{benzenelengths}
Bond lengths ($r$) for the two lowest $^2 B_{2g}$
and $^2 B_{3g}$ states of the benzene radical cation. The angles $\alpha$ 
 are expressed
in degrees, the lengths in $a_0$. The carbon sites are numerated from 1 to 6.}
\begin{ruledtabular}
\begin{tabular}{l d d r }
& \makebox[0pt][c]{$^2 B_{2g}$} & \makebox[0pt][c]{$^2 B_{3g}$}  
& Computational method \\
& \makebox[0pt][c]{acute} & \makebox[0pt][c]{obtuse} 
& \\
\hline
$r(C_1-C_2)$  & 2.616    &     2.694    &  B3LYP/cc-pVTZ \footnotemark[1]  \\
              & 2.649    &   2.725    & BLYP/6-31G* \footnotemark[2] \\
              & 2.659(1) &   2.733(4) & SR-VMC \footnotemark[3] \\
$r(C_2-C_3)$  & 2.735    &   2.579    &    B3LYP/cc-pVTZ \footnotemark[1]  \\
              & 2.766    &  2.615    &BLYP/6-31G* \footnotemark[2] \\
              & 2.764(2) & 2.628(4) & SR-VMC \footnotemark[3] \\
$\alpha(C_6 C_1 C_2) $ & 118.4 & 121.6   &   B3LYP/cc-pVTZ \footnotemark[1] \\
              & 118.5    &  121.5    & BLYP/6-31G* \footnotemark[2] \\
              & 118.95(6) & 121.29(17) & SR-VMC \footnotemark[3] \\
\end{tabular}
\footnotetext[1]{Ref.~\onlinecite{deleuze}}
\footnotetext[2]{Ref.~\onlinecite{klaus}}
\footnotetext[3]{This work}
\end{ruledtabular}
\end{table*}

\newpage

\begin{table*}[!hbp]
\caption{\label{ionized}
Total energies for the 
$^2 B_{2g}$ and $^2 B_{3g}$ states of the benzene radical cation 
after the geometry relaxation. A comparison with a BLYP/6-31G* and 
SVWN/6-31G*  all-electron
calculation (Ref.~\onlinecite{klaus}) is reported.}
\begin{ruledtabular}
\begin{tabular}{l d d d d}
 & \makebox[0pt][c]{VMC} & \makebox[0pt][c]{DMC} &
  \makebox[0pt][c]{BLYP/6-31G*} & \makebox[0pt][c]{SVWN/6-31G*} \\
\hline
$^2 B_{2g}$ & -231.4834(15) & -231.816(3) & -231.815495 & -230.547931 \\
$^2 B_{3g}$ & -231.4826(14) & -231.812(3) & -231.815538 & -230.547751 \\
\end{tabular}
\end{ruledtabular}
\end{table*}

\newpage

\begin{table*}[!hbp]
\caption{\label{adiabatic}
Adiabatic ionization potential of the benzene molecule; 
our estimate is done for the $^2 B_{3g}$ relaxed geometries 
of the benzene radical cation, with an inclusion of the zero point motion
correction between the $^2 B_{3g}$ state and the $^1 A_{1g}$ neutral molecule
ground state, calculated in Ref.~\onlinecite{deleuze} at the B3LYP/6-31G*
 level.}
\begin{ruledtabular}
\begin{tabular}{l d d d d}
& \makebox[0pt][c]{VMC  \footnotemark[1]}
& \makebox[0pt][c]{DMC  \footnotemark[1]}
& \makebox[0pt][c]{CCSD(T)/cc-pV$\infty$Z \footnotemark[2]} &
  \makebox[0pt][c]{experiment \footnotemark[3]} \\
\hline
AIP              &   8.86(6)   &    9.36(8)  & 9.29(4) &  \\
$\Delta ZPE_{ad}$ & -0.074 &  -0.074 & -0.074  & \\
\textbf{best estimate} & 8.79(6) & 9.29(8) & 9.22(4) & 9.2437(8) \\
\end{tabular}
\footnotetext[1]{This work}
\footnotetext[2]{Ref.~\onlinecite{deleuze}}
\footnotetext[3]{Ref.~\onlinecite{ion}}

\end{ruledtabular}
\end{table*}
\newpage

\begin{table*}[!hbp]
{\tiny
\caption{\label{lambda1}
Matrix of the $\lambda$ coefficients of the geminal function expansion in the
pairing determinant for the $Li_2$ molecule. The matrix is symmetric to have a
spin singlet, therefore we show only the upper part of it.} 
\begin{ruledtabular}
\begin{tabular}{l | *{12}{c}}
& \makebox[0pt][c]{$1s_a$}
& \makebox[0pt][c]{$2s_a$}
& \makebox[0pt][c]{$2pz_a$} 
& \makebox[0pt][c]{$2px_a$}
& \makebox[0pt][c]{$2py_a$}
& \makebox[0pt][c]{$3s_a$}
& \makebox[0pt][c]{$1s_b$}
& \makebox[0pt][c]{$2s_b$}
& \makebox[0pt][c]{$2pz_b$} 
& \makebox[0pt][c]{$2px_b$}
& \makebox[0pt][c]{$2py_b$}
& \makebox[0pt][c]{$3s_b$}
 \\
\hline
$1s_a$      &   1   &   $-2.162 \, 10^{-3}$  & $ -6.838 \, 10^{-3}$ & 0 & 0 & $-2.877 \, 10^{-3}$ & 0 & 0 &0  &0 &0 &0  \\
$2s_a$      &   -   &   $6.22 \, 10^{-4}$  &  $-1.601 \, 10^{-4} $   & 0 & 0 & $ 2.031 \, 10^{-3}$ & 0 &$6.79 \, 10^{-3}$   & $-4.67 \, 10^{-4} $ &0 & 0 & $1.790 \, 10^{-3}$ \\
$2pz_a$      &   -   &  - & $-4.25 \, 10^{-4}$  & 0 & 0 & $-1.132 \, 10 ^{-3} $  & 0 & $-4.67 \, 10^{-4} $   &  $3.211 \, 10^{-4}$  &0 & 0 &$7.67\, 10^{-4}$  \\
$2px_a$      &   -   &  - & -  & $-1.351 \, 10^{-3} $& 0 & 0 & 0 & 0 & 0  & $-1.173 \, 10^{-3}$ & 0 & 0  \\
$2py_a$      &   -   &  - & -  & - & $-1.351 \, 10^{-3} $ & 0 & 0 & 0 & 0 &0 & $-1.173 \, 10^{-3}$ & 0  \\
$3s_a$      &   -   &  - & -  & - & - & $-1.541\,10^{-3}$ & 0 & $1.790 \, 10^{-3}$ & $7.67 \,10^{-4}$ &0 & 0 & $-8.91 \,10^{-4}$  \\
$1s_b$      &   -   &  - & -  & - & - & - & 1 &  $-2.162 \, 10^{-3}$ &  $ -6.838 \, 10^{-3}$ &0 & 0 & $-2.877 \, 10^{-3}$  \\
$2s_b$      &   -   &  - & -  & - & - & - & - & $6.22 \, 10^{-4}$  &  $-1.601 \, 10^{-4} $   &0 & 0 &  $1.790 \, 10^{-3}$  \\
$2pz_b$      &   -   &  - & -  & - & - & - & - & - &  $-4.25 \, 10^{-4}$ &0 & 0 & $-1.132 \, 10 ^{-3} $  \\
$2px_b$      &   -   &  - & -  & - & - & - & - & - & - &  $-1.351 \, 10^{-3} $   & 0 & 0  \\
$2py_b$      &   -   &  - & -  & - & - & - & - & - &  -  & - & $-1.351 \, 10^{-3} $  & 0   \\
$3s_b$      &   -   &  - & -  & - & - & - & - & - &  -  & - & -  & $-1.541\,10^{-3}$   \\
\end{tabular}
\end{ruledtabular}
}
\end{table*}

\newpage

\begin{table*}[!hbp]
\caption{\label{lambda2}
Matrix of the $\lambda$ coefficients of the pairing function expansion in the
three body Jastrow for the $Li_2$ molecule. 
As in the previous table only the upper part is reported .}
\begin{ruledtabular}
\begin{tabular}{l | *{8}{c}}
& \makebox[0pt][c]{$sG_a$}
& \makebox[0pt][c]{$pGx_a$}
& \makebox[0pt][c]{$pGy_a$} 
& \makebox[0pt][c]{$pGz_a$}
& \makebox[0pt][c]{$sG_b$}
& \makebox[0pt][c]{$pGx_b$}
& \makebox[0pt][c]{$pGy_b$} 
& \makebox[0pt][c]{$pGz_b$}
 \\
\hline
$sG_a$ & $-0.2427$ &  0 & 0  & $-2.713 \, 10^{-4}$ & $-5.136 \, 10^{-4}$ & 0 & 0 & $-1.202 \, 10^{-5}$   \\
$pGx_a$ & - &  $-0.1772$  & 0  & 0 & 0 & $-7.997 \, 10^{-3}$ & 0 & 0  \\
$pGy_a$ & - &   -   & $-0.1772 $  & 0 & 0 & 0 & $-7.997 \, 10^{-3}$ & 0  \\
$pGz_a$ & - &   -   & -  & $1.027 \, 10^{-2}$ & $1.202 \, 10^{-5}$ & 0 & 0 & $-8.749 \, 10^{-3}$  \\
$sG_b$ & - &   -   & -  & - & $-0.2427$ & 0 & 0 & $-2.713 \, 10^{-4}$  \\
$pGx_b$ & - &   -   & -  & - & - & $-0.1772$ & 0 & 0  \\
$pGy_b$ & - &   -   & -  & - & - & - & $-0.1772$ & 0  \\
$pGz_b$ & - &   -   & -  & - & - & -  & - &  $1.027 \, 10^{-2}$  \\
\end{tabular} 
\end{ruledtabular}
\end{table*}

\newpage

\begin{table}[!hbp]
\caption{\label{parli2}
Orbital basis set parameters used for the $Li_2$ molecule. Since the molecule
is homonuclear the parameters of the atom $b$ are the same as the atom $a$.} 
\begin{ruledtabular}
\begin{tabular}{l | *{3}{d}}
& \makebox[0pt][c]{$z_1$}
& \makebox[0pt][c]{$z_2$}
& \makebox[0pt][c]{$p$} 
 \\
\hline
$\phi_{1s_a}$   &  2.4485 & 4.2891 & 0.4278  \\
$\phi_{2s_a}$   &  0.5421 & 1.4143 & -1.5500 \\
$\phi_{2px_a}$  &  0.6880 & - & - \\
$\phi_{2py_a}$  &  0.6880 & - & - \\
$\phi_{2pz_a}$  &  1.0528 & - & - \\
$\phi_{3s_a}$   &  0.6386 & - & - \\
$\phi_{sG_a}$   &  1.4356 & - &-0.2044  \\
$\phi_{pGx_a}$  &  0.7969 &4.4217   &-1.2689 \\
$\phi_{pGy_a}$  &  0.7969 &4.4217   &-1.2689 \\
$\phi_{pGz_a}$  &  8.980~\mbox{$10^{-3}$}  &-0.1924 &0.3229  \\
\end{tabular} 
\end{ruledtabular}
\end{table}


\begin{thebibliography}{99}

\bibitem{heitler} W. Heitler and F. London, Z. Physik \textbf{44}, 455 (1927).
\bibitem{barbiellini} B. Barbiellini, J. Phys. Chem. Solids \textbf{61}, 341
  (2000). 
\bibitem{evangelisti} S. Evangelisti, G. L. Bendazzoli, R. Ansaloni, F. Duri,
  E. Rossi, Chem. Phys. Lett. \textbf{252}, 437 (1996).
\bibitem{libro} T. D. Crawford and H. F. Schaefer III, 
in \emph{Reviews in  Computational Chemistry}, edited by K. B. Lipkowitz and D. B. Boyd
 (VCH Publishers, New York, 1991), Vol. 14, pp. 33-136. \emph{An introduction
   to Coupled Cluster Theory for Computational Chemists}. 
\bibitem{ferro} L. Noodleman, T. Lovell, T. Liu, F. Himo, R. A. Torres,
  Curr. Opin. Chem. Biol. \textbf{6}, 259 (2002)
\bibitem{vdW} W. Kohn, Y. Meir, and D. E. Makarov,
  Phys. Rev. Lett. \textbf{80}, 4153 (1998); M. Lein, J. F. Dobson,
  E. K. U. Gross, J. Comp. Chem. \textbf{20}, 12 (1999); 
 H. Rydberg, M. Dion, N. Jacobson, E. Schr\"oder, P. Hyldgaard, S. I. Simak,
  D. C. Langreth, and B. I. Lundqvist, Phys. Rev. Lett. \textbf{91}, 126402
  (2003).  
\bibitem{fahy} S. Fahy, X. W. Wang and S. G. Louie, Phys. Rev. B
  \textbf{42}, 3503 (1990).
\bibitem{umrigarint} C. J. Umrigar in \emph{Quantum Monte Carlo Methods in
Physics and Chemistry}, Proceedings of the NATO Advanced Study Institute, edited
by M. P. Nightingale and C. J. Umrigar (Kluwer, Dordrecht, 1998), page 129.
\bibitem{dmc} P. J. Reynolds and D. M. Ceperley, B. J. Alder, W. A. Lester,
  J. Chem. Phys. \textbf{77}, 5593 (1982); 
L. Mit\'a\v{s},
  E. L. Shirley, D. M. Ceperley, J. Chem. Phys. \textbf{95}, 3467 (1991);
C. J. Umrigar, M. P. Nightingale,
  K. J. Runge, J. Chem. Phys. \textbf{99}, 2865 (1993).
\bibitem{casula} M. Casula and S. Sorella, J. Chem. Phys. \textbf{119}, 
6500 (2003).
\bibitem{foulkes} W. M. C. Foulkes, L. Mitas, R. J. Needs and G. Rajagopal,
  Rev. Mod. Phys. \textbf{73}, 33 (2001)
\bibitem{coleman} A. J. Coleman, J. Math. Phys. \textbf{13}, 214 (1972).
\bibitem{cep2body} D. Ceperley, Phys. Rev. B \textbf{18}, 3126 (1978).
\bibitem{umrigar3body} C. J. Umrigar, K. G. Wilson and J.W. Wilkins, Phys. Rev. Lett. \textbf{60}, 1719 (1988).
\bibitem{prendergast} D. Prendergast, D. Bevan and S. Fahy, Phys. Rev. B \textbf{66}, 155104 (2002).

\bibitem{krotscheck} E. Krotscheck, W. Kohn and G. X. Qian, Phys. Rev. B
  \textbf{32} 5693 (1985).
\bibitem{sorella} S. Sorella, Phys. Rev. B \textbf{64},
024512 (2001).
\bibitem{night} M. P. Nightingale and V. Melik-Alaverdian, Phys. Rev. Lett. \textbf{87}, 43401 (2001).
\bibitem{filippi2} F. Schaultz and C. Filippi, J. Chem. Phys. \textbf{120},10931 (2004).
\bibitem{zeroforce} R. Assaraf and M. Caffarel, Phys. Rev. Lett. \textbf{83},
4682 (1999); R. Assaraf and M. Caffarel, J. Chem. Phys. \textbf{113},
4028 (2000).
\bibitem{penalty} D. M. Ceperley and M. Dewing, J. Chem. Phys. \textbf{110},
9812 (1999).
\bibitem{correlated} Z. Sun and W. A. Lester, J. Chem. Phys. \textbf{97},
7585 (1992).
\bibitem{umrigar} C. J. Umrigar, Int. J. Quantum Chem. Symp. \textbf{23},
217 (1989).
\bibitem{filippi} C. Filippi and C. J. Umrigar, Phys. Rev. B \textbf{61},
R16291 (2000).
\bibitem{car} R. Car and M. Parrinello, Phys. Rev. Lett. \textbf{55},
2471 (1985).
\bibitem{tanaka} S. Tanaka, J. Chem. Phys. \textbf{100},
7416 (1994).
\bibitem{mella} M. Casalegno, M. Mella and A. M. Rappe,
  J. Chem. Phys. \textbf{118}, 7193 (2003).
\bibitem{moroni} S. De Palo, S. Moroni, S. Baroni, cond-mat/0111486
\bibitem{filippimol} C. Filippi and C. J. Umrigar,
  J. Chem. Phys. \textbf{105}, 213 (1996).
\bibitem{assaraf} R. Assaraf and M. Caffarel, J. Chem. Phys. \textbf{119}, 
10536 (2003).
\bibitem{varianza} M. Snajdr. S.M. Rothstein J. Chem. Phys. \textbf{112}, 4935 (2000);
 D. Bressanini, G. Morosi and M. Mella, J. Chem. Phys. \textbf{116}, 5345 (2002); F.J. Galvez, E. Buendia, A. Sarsa, J. Chem. Phys. \textbf{115}, 1166 (2001).
\bibitem{grossman} J. C. Grossman, J. Chem. Phys. \textbf{117}, 
1434 (2002).
\bibitem{huang} H. Huang and Z. Cao, J. Chem. Phys. \textbf{104},
200 (1996).
\bibitem{garmer} D.R.Garmer and J.B. Anderson, J. Chem. Phys. \textbf{86},
  4025 (1987).
\bibitem{lu} Shih-I Lu, J. Chem. Phys. \textbf{118}, 9528 (2003).
\bibitem{luchow} A. L\" uchow and J. B. Anderson, D. Feller,
  J. Chem. Phys. \textbf{106}, 7706 (1997).
\bibitem{martin} J. M. L. Martin, Chem. Phys. Lett. \textbf{303}, 399 (1999).
\bibitem{roeggen} I. R\o eggen, J. Chem. Phys. \textbf{79}, 5520 (1983).
\bibitem{roeggen2} I. R\o eggen and J. Alml\" of, Int. J. Quantum
  Chem. \textbf{60}, 453 (1996).
 \bibitem{gutz} P. Horsch \prb {\bf 24} 7351 (1981); D. Baeriswyl and K. Maki
\prb {\bf 31}, 6633 (1985). 
\bibitem{pauling} L. Pauling, \emph{The nature of the chemical bond}, Third
  edition (Cornell University Press, Ithaca, New York), page 204.
\bibitem{feller2} D. Feller and D. A. Dixon, J. Phys. Chem. A \textbf{104},
  3048 (2000). 
\bibitem{srinivasan} Srinivasan Parthiban and J.M.L. Martin,
  J. Chem. Phys. \textbf{115}, 2051 (2001).
\bibitem{klaus} K. Muller-Dethlefs, J. B. Peel, J. Chem. Phys. \textbf{111},
10550 (1999).
\bibitem{deleuze} M. S. Deleuze, L. Claes, E. S. Kryachko, and
  J.P. Fran\c cois, J. Chem. Phys. \textbf{119}, 3106 (2003).
\bibitem{fahy2} C. Filippi and S. Fahy, J. Chem. Phys. \textbf{112}, 3523
(2000).
\bibitem{bcsagp} B. Weiner and O. Goscinski, \pra {\bf 22}, 2374 (1980).
\bibitem{hubbard} see e.g. 
A. Paramekanti, M. Randeria and N. Trivedi, \prl {\bf 87}, 217002 (2001);
 T. Nakanishi, K. Yamaji and T. Yanagisawa, J. Phys. Soc. Jpn. {\bf 66},
   294 (1997);  H. Yokoyama, Y. Tanaka, M. Ogata and H. Tsuchiura, cond-mat/0308264.
\bibitem{capello} M. Capello, F. Becca, M. Fabrizio, S. Sorella and
  E. Tosatti, cond-mat/0403430.
\bibitem{exact} S. J. Chakravorty, S. R. Gwaltney, E. R. Davidson,
F. A. Parpia, and C. F. Fischer, Phys. Rev. A \textbf{47}, 3649 (1993).
\bibitem{feller} D. Feller, C. M. Boyle, and E. R. Davidson,
  J. Chem. Phys. \textbf{86}, 3424 (1987).
\bibitem{ermler} W. C. Ermler and C. W. Kern, J. Chem. Phys. \textbf{58}, 
3458 (1973).
\bibitem{ion} \emph{Ion Energetics Data} in \emph{NIST Chemistry Webbook, NIST
  Sandard Reference Database} Number 69, edited by P. J. Linstrom and
  W. G. Mallard (National Institute of Standards and Technology, Gaithersburg,
  MD, 2001) (http://webbook.nist.gov).
\end{thebibliography}
\end{document}